\documentclass[11pt,a4paper]{article}
\pdfoutput=1

\usepackage{jcappub}

\usepackage{array}
\usepackage{accents}
\usepackage{nicefrac}
\usepackage{setspace}
\usepackage{booktabs}
\usepackage{dcolumn}
\usepackage{multirow}
\usepackage{wasysym}
\usepackage{subfigure}

\renewcommand{\d}{\mathrm{d}}

\renewcommand{\l}{\ell}

\title{The neutrino signal at HALO: learning about the primary supernova neutrino fluxes and neutrino properties}

\author{Daavid V\"a\"an\"anen,}
\author{Cristina Volpe}

\affiliation{Institut de Physique Nucl\'eaire, \\
F-91406 Orsay cedex, CNRS/IN2P3 and University of Paris-XI, France}

\emailAdd{vaananen@ipno.in2p3.fr}
\emailAdd{volpe@ipno.in2p3.fr}

\abstract{Core-collapse supernova neutrinos undergo a variety of phenomena when they travel from the high neutrino density region and large matter densities to the Earth. We perform analytical calculations of the supernova neutrino fluxes including collective effects due to the neutrino-neutrino interactions, the Mikheev-Smirnov-Wolfenstein (MSW) effect due to the neutrino interactions with the background matter and decoherence of the wave packets as they propagate in space. We predict the numbers of one- and two-neutron charged and neutral-current electron-neutrino scattering on lead events. We show that, due to the energy thresholds, the ratios of one- to two-neutron events are sensitive to the pinching parameters of neutrino fluxes at the neutrinosphere, almost independently of the presently unknown neutrino properties. Besides, such events have an interesting sensitivity to the spectral split features that depend upon the presence/absence of energy equipartition among neutrino flavors. Our calculations show that a lead-based observatory like the Helium And Lead Observatory (HALO) has the potential to pin down important characteristics of the neutrino fluxes at the neutrinosphere, and provide us with information on the neutrino transport in the supernova core.}
\notoc

\begin{document}
\maketitle

\flushbottom

\section{Introduction}
Parallel to the developments occurring in core-collapse supernova simulations \cite{Mezzacappa:2005ju}, the degree of complexity in understanding of neutrino flavor conversion in dense matter has enormously increased in the last years. 
It was pointed out some time ago \cite{Pantaleon1992eq,Samuel:1993uw} that 
the neutrino-neutrino interaction could trigger new mechanisms for flavor conversion 
since the neutrino number density within the supernova is very large.
The ongoing investigations focus on unraveling these flavor conversion phenomena,
on the underlying mechanisms and/or general properties of the associated non-linear Hamiltonian
and the phenomenological impact. It is now understood that  
in the inner region, near the neutrinospheres, flavor conversion is first frozen by synchronization
 due to the strength of the neutrino-neutrino interactions \cite{Duan:2005cp,Duan:2006an}. 
Then bipolar oscillation occur \cite{Duan:2006an,Hannestad:2006nj}, ending up in spectral splits 
and swaps \cite{Duan:2005cp,Duan:2006an,Raffelt:2007xt,Dasgupta:2007ws,Dasgupta:2009mg,Dasgupta:2010cd}.
Such collective effects can be suppressed by decoherence coming from large matter 
densities, as pointed out in e.g. \cite{EstebanPretel:2007ec}), and 
recently investigated using one-dimensional matter density profiles 
from supernova simulations \cite{Chakraborty:2011gd}. 
The presence of invariants of the Hamiltonian describing the neutrino evolution has been
identified in several works, e.g. \cite{Raffelt:2007xt,Duan:2008fd}. 
This aspect has been further investigated in a recent interesting work \cite{Pehlivan:2011hp}
where it has been shown the presence of invariants not only for the effective one-body mean-field description 
but for the many-body Hamiltonian (when the matter term is not included). 
The impact  
of the matter phase (and its derivative) -- coming from the presence of complex 
off-diagonal contributions from the neutrino-neutrino interaction term -- on the neutrino evolution in the 'matter' basis Hamiltonian is pointed out in \cite{Galais:2011jh}. In particular, 
the start of bipolar oscillations is shown to be associated with a rapid growth of the phase derivative. 
On the other hand bipolar oscillations 
can be understood as the motion of a flavor pendulum \cite{Duan:2005cp} and gyroscopic pendulum \cite{Hannestad:2006nj,Wu:2011yi}.
By building up a self-consistent adiabatic solution in a comoving frame, in \cite{Duan:2007mv, Raffelt:2007xt} the authors discuss that the spectral split is a regular precession phenomenon. In \cite{Galais:2011gh} the spectral split phenomenon is put in relation with a magnetic resonance phenomenon. 
By realizing a full numerical calculation in two-flavors and the single-angle approximation it is shown 
that the neutrino energies and the location where the magnetic resonance condition is satisfied are
the same as where the spectral split occurs. Putting the spectral split in connection with the magnetic resonance phenomenon points out the way to calculate the resonance condition in the flavor basis directly, without going to the comoving frame.
In \cite{Wu:2011yi} both precession and nutation driven resonance conditions governing the spectral split are discussed.
Besides several works have started investigating the phenomenological implications of collective effects.
The possibility to pin down the neutrino hierarchy and third neutrino mixing angle using 
shadowing from Earth matter effects is studied in \cite{Dasgupta:2008my} or the
positron time signal associated with inverse beta decay in \cite{Gava:2009pj}. 
The impact on the diffuse supernova neutrino background is studied in
\cite{Chakraborty:2010sz,Galais:2009wi}.
While important progress has been made in the last years,
several aspects still need a full understanding as well as finally assessing
the phenomenological impact of collective 
effects in a future core-collapse supernova signal.

When the neutrino interactions with ordinary matter starts dominating, the 
Mikheev-Smirnov-Wolfenstein (MSW) effect induces efficient flavor conversion 
if the matter density profile is adiabatic \cite{W,MS}. However the arrival of the (front and reverse) 
shocks in the outer layers of the star can leave an imprint on the neutrino signal 
\cite{Schirato:2002tg,Fogli:2004ff}, render the conversion non-adiabatic, and induce 
multiple resonances and phase effects \cite{Dasgupta:2005wn,Kneller:2007kg}. A full 
calculation of the neutrino amplitudes up to the star surface, using matter density 
profiles from hydrodynamical calculations, has shown that no extra interference 
effects occur from the combination of the two regions in iron core-collapse supernovae 
\cite{Gava:2009pj}. At the end of their trip, decoherence~\cite{decoh} and Earth matter effects~\cite{Akhmedov:1998ui} impact the neutrino wave packets before reaching an observatory on Earth.  

The details of the neutrino spectra at the surface of the star depend upon unknown neutrino parameters -- in particular the mass hierarchy and the third neutrino mixing angle -- and upon the "primary" neutrino fluxes (at the neutrinosphere). In addition, another important unknown property is the possible existence of CP violation in the leptonic sector. Possible effects coming from leptonic CP violation in media are being explored. First studied in \cite{Akhmedov:2002zj}, it has recently been shown that the Dirac CP violating phase can modify the neutrino fluxes in media, both in the supernova environment \cite{Balantekin:2007es,Gava:2008rp} and in the Early Universe at the epoch of Big-Bang nucleosynthesis \cite{Gava:2010kz}. 
Such effects arise for example because of the different interaction with matter that muon and tau neutrinos have, due both to radiative corrections within the Standard model and to physics beyond the Standard model (e.g. Flavor-Changing Neutral currents). 
The features of the neutrino fluxes at the neutrinosphere encode information on various aspects of the supernova dynamics, including the microscopic processes determining the neutrino transport within the supernova core and the equation of state of the neutron star (for the late time cooling phase). For example, simulations show that, in the case of very massive stars leading to failed supernovae and black holes, larger average neutrino energies can be achieved compared to the case of supernovae turning to neutron stars \cite{Sumiyoshi:2007pp}. 

The neutrino spectra appear to be rather well fitted using Fermi-Dirac or power-law distributions \cite{Keil:2002in}. Beside the average energy these spectra are characterized by the second moment of the distribution related to the pinching parameter. Supernova simulations suggest non-zero values for this parameter~\cite{Myra:1990tt,Keil:2002in}. While a hierarchy of average neutrino energies is expected due to the difference of the electron neutrino, anti-neutrino and muon and tau (anti)neutrino cross sections with matter from the standard model at tree level, the values of average energies themselves are subject to variations depending on the supernova simulations (see \cite{Keil:2002in} and references therein). 
Furthermore the presence of an equipartition of the gravitational energy emitted during the supernova explosion among the flavors is one of the open issues. Indeed, after the prompt $\nu_e$ burst, 
the electron-type neutrino luminosities 
can deviate from the non-electron-type (anti)neutrino luminosities 
within a factor of two,
depending on the simulations and the phase of the explosion (see e.g. table 7.3 of ref.~\cite{Keil:2003sw}, table 6 in~\cite{Keil:2002in} and~\cite{Fischer:2009af}). This also produces qualitatively different neutrino spectra after collective effects depending on the primary neutrino flux parameters~\cite{Choubey:2010up}. 

Several running neutrino detectors are waiting for the next core-collapse supernova explosion, while new large size observatories based on various technologies are under study \cite{Autiero:2007zj}. The latter would look for proton decay, CP violation \cite{Volpe:2006in} and (extra-)galactic supernova neutrinos having, in particular, a discovery reach for the diffuse supernova neutrino background \cite{DSNB}. While most of such observatories are sensitive to electron anti-neutrinos through scattering on protons, the detection of electron neutrinos exploits scattering on nuclei. 

A new detector is currently under construction at SNOLAB: the Helium And Lead Observatory (HALO) \cite{Duba:2008zz}. This dedicated supernova neutrino detector is able to observe the neutrons emitted from electron-neutrino scattering on lead from charged- and neutral-current events \cite{Duba:2008zz}. While the idea was first proposed a long time ago \cite{Cline:1993rx} different possible realizations have been discussed in the literature \cite{Smith:1997td,Zach:2002is,Hargrove:1996zv,Elliott:2000su}. In \cite{Fuller:1998kb} it is shown that, among different possible nuclei as neutrino targets, measuring one- and two-neutron events on lead is particularly attractive to extract information on the supernova neutrino temperatures. A subsequent work \cite{Engel:2002hg} has shown that these detection channels are very sensitive to large/small values of the still unknown third neutrino mixing angle $\theta_{13}$ and that the possible detection of the outgoing electron energy in the detector could help extracting information on the pinching parameters of the primary neutrino fluxes. 

The goal of this paper is to investigate the information that can be extracted on the 
characteristics of the neutrino spectra at the neutrinosphere with a detector like HALO, 
in spite of the existing uncertainties coming from the unknown neutrino properties and 
from the different supernova simulations. With this aim we perform new predictions of the 
charged- and neutral-current supernova neutrino signal expected in HALO. 
The lead-based supernova neutrino detector currently built is made of solid lead so that no electron can be measured,
at variance with some of the detector designs previously considered.
Having such a design in mind, 
we perform a detailed analysis of the information that can be extracted from the measurement 
of one- and two-neutron events, if an iron core-collapse supernova explodes in our galaxy. 
Our calculations include for the first time the 
collective effects due to the neutrino-neutrino interactions, the 
MSW effect in the outer layers of the star and the decoherence introduced by the traveling up to the Earth. 

The manuscript is structured as follows. In Section II we present the formalism and our analytical treatment of the neutrino propagation, from the neutrinosphere up to a detector on the Earth. Section III describes the numerical results on the neutrino fluxes and the expected number of events in a lead-based observatory like HALO. The potential to unravel information on the neutrino spectra is analyzed.  
Section IV is a conclusion.

\section{The neutrino evolution from production to detection: the formalism}
\noindent
In this work we will follow the neutrino flavor evolution from the neutrinosphere up to the Earth. We evolve the neutrino probabilities. 
The assumption of propagating the probabilities and not the amplitudes is sufficient for our purposes since we do not aim at capturing any phase effects, like those arising from the Dirac CP violation phase \cite{Balantekin:2007es,Gava:2008rp} or from the presence of multiple resonances \cite{Dasgupta:2005wn,Kneller:2007kg,Gava:2009pj}. Possible effects arising from the radiative corrections inducing the $V_{\mu\tau}$ potential will not be considered here. Note that these might arise in the case of very large matter densities \cite{Pastor:2008zz} or of non-standard contributions from SUSY models \cite{Gava:2009gt}. 

Therefore, non-electron-type (anti)neutrino fluxes are assumed to be equal and it is convenient to work in the $\nu_e, \nu_{x}, \nu_{y}$ basis, which corresponds to mass eigenstate basis at high background matter densities  close to the neutrino production region~\cite{Kuo:1986sk}. This basis is related to the $\nu_e,  \nu_{\mu}, \nu_{\tau}$ basis through a rotation along the second mixing angle $\theta_{23}$: 
\begin{equation}
 \label{eq:R23}
 \begin{split}
 \nu_{x} &= \cos\theta_{23} \nu_{\mu} - \sin\theta_{23}\nu_{\tau} \ , \\
 \nu_{y} &= \sin\theta_{23} \nu_{\mu} + \cos\theta_{23}\nu_{\tau} \ ,
 \end{split}
\end{equation}
such that \textit{x}- and \textit{y}-type (anti)neutrino fluxes are equal to other non-electron-type fluxes. 
Following the above assumptions the observable final flavor neutrino fluxes at the Earth $F = \left( F_{\nu_e}, F_{\nu_{\mu}}, F_{\nu_{\tau}} \right)^T$ can be computed as follows: 
\begin{equation}
	\label{e:SNF}
	F = A P_{\rm{MSW}} P_{\nu\nu}  F^0 \ , 
\quad F^0 =
	\begin{pmatrix}
		F^0_{\nu_e} \\
		F^0_{\nu_{x}} \\
		F^0_{\nu_{y}}
	\end{pmatrix} \ ,
\end{equation}
where $F^{0}_{\nu_e}, F^{0}_{\nu_x}, F^{0}_{\nu_y}$ are the primary neutrino fluxes at the neutrinosphere, $P_{\nu\nu}$ is the neutrino-neutrino interaction swapping probability matrix, $P_{\rm MSW}$ is the MSW conversion probability matrix and A takes into account the decoherence of the neutrino wave packets after neutrinos exit the surface. In this notation it is understood that $F_{\nu_{\l}} = 1/(4\pi r^2) F_{\nu_{\l}}\ (\l = e, \mu, \tau)$, where $r$ is the distance to the supernova.
 
Our eq.~\eqref{e:SNF} is assuming factorized dynamics. This hypothesis is based on the fact that, within the MSW region, the high and low resonance are well separated in iron core-collapse supernova~\cite{Dighe:1999bi}. In \cite{Gava:2009pj} a complete propagation of the neutrino amplitudes from the neutrinosphere to the star surface has been performed and has shown that it is indeed a good approximation to factorize the dynamics between the inner region, where the neutrino-neutrino interaction effects are important, and the outer layers, where the MSW resonances occur. 

Let us give the explicit expression of each matrix in eq.~\eqref{e:SNF}. For the collective flavor conversion effects in the neutrino-neutrino interaction region, it is sufficient to consider the following swapping probability matrix: 
\begin{equation}
	\label{eq:Pnunu}
	P_{\nu\nu} =
	\begin{pmatrix}
		P_{\l\l} & P_{ex} & P_{ey}	\\
		P_{ex} & 1 - P_{ex} & 0	\\
		P_{ey} & 0 &	1 - P_{ey}
	\end{pmatrix} \ ,
\end{equation}
where $P_{\l\l} = P(\nu_{\l} \rightarrow \nu_{\l}) = 1 - P_{ex} - P_{ey} \ (\l = e, x, y)$ corresponds to neutrino survival probability, while $P_{ex} = P(\nu_e \leftrightarrow \nu_x)$ and $P_{ey} = P(\nu_e \leftrightarrow \nu_y)$ describe the electron neutrino swapping with {\itshape x}- and {\itshape y}-type neutrinos, respectively, such that $P_{\l\l} + P_{ex} + P_{ey} = 1$. For antineutrinos one can use the same form as above by replacing $\nu_{\l}$ with $\bar{\nu}_{\l}$.
 
After the collective effects have ceased, the neutrinos enter the MSW region, in which the background matter can significantly modify neutrino flavor evolution~\cite{W,MS}. One can construct a matrix consisting of all the possible probabilities that original flavor (anti)neutrino state $\nu_{\l}$ arrives at the surface of the star as a mass eigenstate $\nu_i$: $P_{\l i} = P(\nu_{\l} \rightarrow \nu_i) \ (\l = e, x, y; i = 1, 2, 3)$. In normal mass hierarchy (NMH) the MSW level crossing probability matrix for neutrinos is given by \cite{Dighe:1999bi}
\begin{equation}
  \begin{split}
	\label{eq:PMSW}
	P_{\rm{MSW}} &\equiv
	\begin{pmatrix}
		P_{e 1} & P_{x 1} & P_{y 1} \\
		P_{e 2} & P_{x 2} & P_{y 2} \\
		P_{e 3} & P_{x 3} & P_{y 3}
	\end{pmatrix} \\ &=
	\begin{pmatrix}
		P_H P_L & 1 - P_L & (1 - P_H) P_L \\
		P_H (1 - P_L) & P_L & (1 - P_H) (1 - P_L) \\
		1 - P_H & 0 & P_H
	\end{pmatrix} \ ,
  \end{split}
\end{equation}
where $P_H$ and $P_L$ are the high and low resonance level-crossing probabilities, respectively. The corresponding matrix in inverted mass hierarchy (IMH) is obtained from above by setting $P_H = 1$. In the case of antineutrinos the crossing probability matrix $\overline{P}_{\rm{MSW}}$ in NMH is a unit matrix since no resonances occur in the antineutrino sector in this case. The antineutrino conversion matrix in IMH is obtained from eq.~\eqref{eq:PMSW} by replacing $P_L$ by $1 - \overline{P}_L$.   
Note that the high resonance crossing probabilities are equal in the neutrino and antineutrino sectors $(\overline{P}_H = P_H)$.

The neutrino flavor evolution depends on how the resonances are crossed. The crossing is completely adiabatic ($P_{L, H} = 0$) if neutrino remains on its original mass eigenstate and completely non-adiabatic ($P_{L, H} = 1$) if it jumps to the other energy eigenstate. The adiabaticity of the resonances in turn depends on neutrino mixing parameters and the matter density profile~\cite{Dighe:1999bi}.
Note that the presence of large matter density fluctuations can break the H- and L-factorization as discussed in \cite{Kneller:2010ky,Kneller:2010sc}. While in the present work we do not include either shock wave or turbulence effects a discussion will be made of how our results might be changed if one implements supernova matter density profiles with a (front and reverse) shock  \cite{Dasgupta:2005wn,Kneller:2007kg,Gava:2009pj} and matter density fluctuations \cite{Fogli:2003dw,Kneller:2010sc}.  

Once the (anti)neutrinos have reached the surface of the star as mass eigenstates $\accentset{(-)}{\nu}_i\ (i = 1, 2, 3)$, they travel up to the Earth and are detected as flavor eigenstates $\accentset{(-)}{\nu}_{\l}\ (\l = e, \mu, \tau)$. For typical distances between a supernova in our Galaxy and the Earth the mass eigenstate wave packets decohere \cite{Dighe:1999bi} so that no extra effect coming from the oscillations occur. This is taken into account by the use of the matrix: 
\begin{equation}
	\label{eq:A}
	A =
	\begin{pmatrix}
		|U_{e1}|^2 & |U_{e2}|^2 & |U_{e3}|^2 \\
		|U_{\mu1}|^2 & |U_{\mu2}|^2 & |U_{\mu3}|^2 \\
		|U_{\tau1}|^2 & |U_{\tau2}|^2 & |U_{\tau3}|^2
	\end{pmatrix} \ ,
\end{equation}
consisting of all the probabilities of a certain flavor neutrino $\nu_{\l}$ being in a given mass eigenstate $\nu_{i}$: $|U_{\l i}|^2 = \left| \langle \nu_{\l} | \nu_i \rangle \right|^2 \ (\l = e, \mu, \tau; i = 1, 2, 3)$. $U_{\l i}$ are the elements of the Pontecorvo-Maki-Nakagawa-Sakata (PMNS) mixing matrix $U_{\rm PMNS}$ which is the unitary rotation matrix relating the flavor and the mass basis. Considering three neutrino flavors, such a matrix depends on three mixing angles $\theta_{12}, \theta_{13}, \theta_{23}$, one Dirac and two (possible) Majorana phases. The standard parametrization of this mixing matrix can be found e.g. in~\cite{Nakamura:2010zzi}. 

Finally, let us discuss the primary neutrino fluxes at the neutrinospheres, $F^0$ in eq.~\eqref{e:SNF}, whose features reflect the underlying neutrino microscopic processes in the neutrino transport. 
According to current supernova simulations, that include neutrino transport by solving the Boltzmann equation for particles without mixings (see e.g. \cite{Rampp:2000ws}), the neutrino fluxes can be rather well parametrized either by Fermi-Dirac or modified power-law distributions. As a consequence, $F^0$ can be defined as 
\begin{equation}
	\label{eq:F0}
	F^0_{\nu}(E_{\nu}) = \frac{L_{\nu}}{\langle E^0_{\nu} \rangle} \phi(E_{\nu}) \ ,
\end{equation}
where $L_{\nu}$ is the neutrino luminosity, $\langle E^0_{\nu} \rangle$ the average neutrino energy and $\phi(E_{\nu})$ the neutrino energy spectrum at the neutrinosphere.

\begin{figure}\label{fig:FD_eta_T}
 \begin{center}
 \includegraphics[width=.6\textwidth]{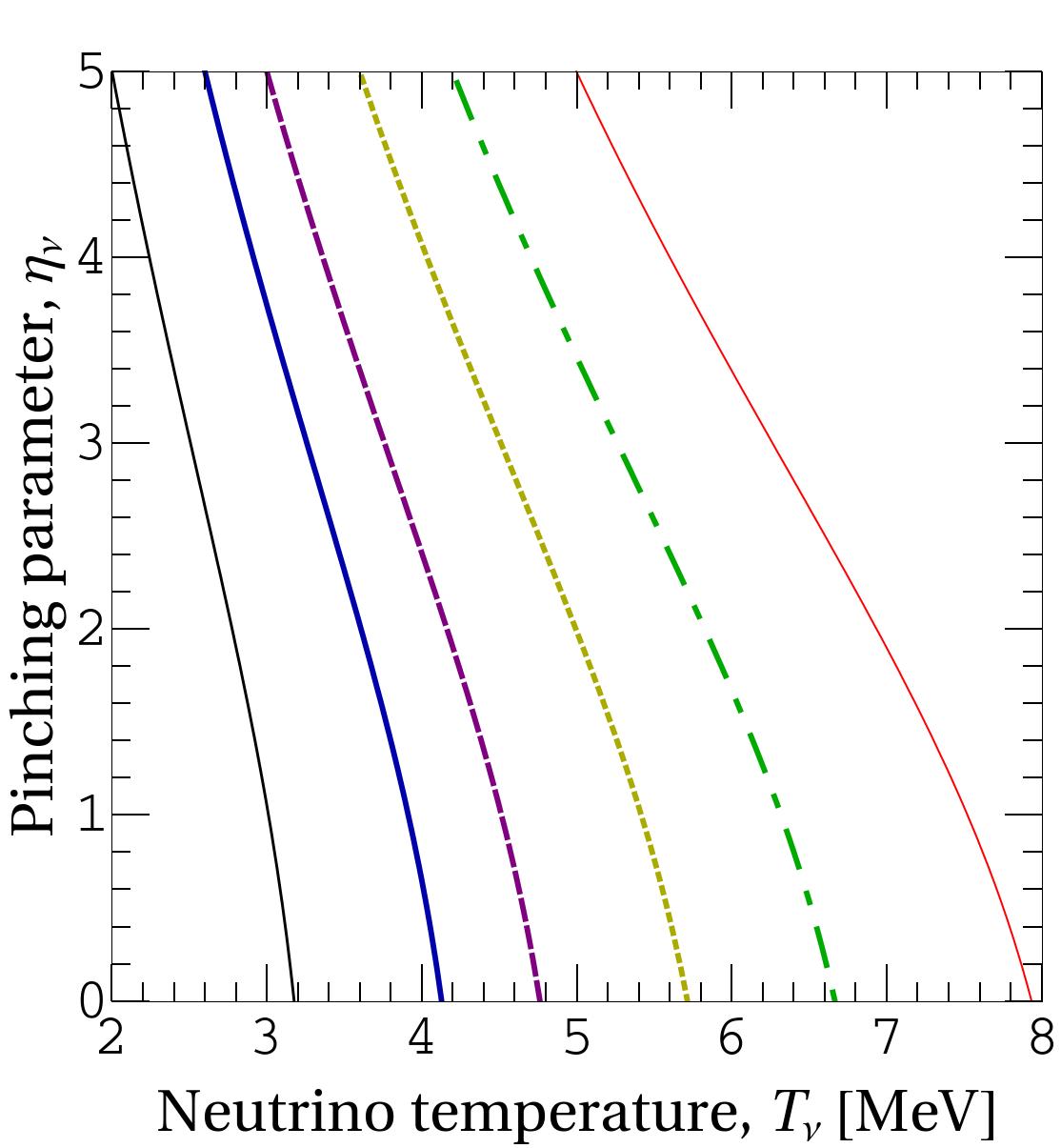}
  \caption{Neutrino degeneracy parameter $\eta_{\nu}$ as a function of neutrino temperature $T_{\nu}$ with different average energies (see eq.~\eqref{eq:Eave}). Neutrino average energies are from left to right: $10, 13, 15, 18, 21, 25\ \mathrm{MeV}$.}
 \end{center}
\end{figure}

For normalized $\left( \int \mathrm{d}E \phi(E) = 1 \right)$ ``pinched'' Fermi-Dirac distribution the energy spectrum can be written as
\begin{equation}
	\label{eq:FDdistr}
	\phi_{FD}(E_{\nu}) = \frac{1}{T_{\nu}^3 F_2(\eta_{\nu})} \frac{E^2_{\nu}}{\mathrm{exp}\left(E_{\nu}/T_{\nu} - \eta_{\nu} \right) + 1} \ ,
\end{equation}
where $T_{\nu}$ is the neutrino temperature, $\eta_{\nu}$ is the degeneracy or pinching parameter and the average neutrino energy can be expressed as
\begin{equation}
	\label{eq:Eave}
	\langle E^0_{\nu} \rangle = T_{\nu}\frac{F_3(\eta_{\nu})}{F_2(\eta_{\nu})} \ .
\end{equation}
The functions $F_2(\eta_{\nu})$ and $F_3(\eta_{\nu})$ are the Fermi integrals:
\begin{equation}
	\label{eq:Fint}
	F_n(\eta_{\nu}) = \int_0^{\infty} \frac{x^n \mathrm{d}x}{\mathrm{exp}\left(x - \eta_{\nu} \right) + 1} \ .
\end{equation}
The relation between $\eta_{\nu}$ and $T_{\nu}$ is plotted in ~\ref{fig:FD_eta_T} using different values for $\langle E_{\nu} \rangle$.

In \cite{Keil:2002in} it is argued that the power-law parametrization is more flexible at describing the high-energy tail of the neutrino spectra in comparison to the above Fermi-Dirac distribution. 
For normalized modified power law or alpha-fit distribution the energy spectrum is given by
\begin{equation}
	\label{eq:PLdistr}
	\phi_{PL}(E_{\nu}) = \mathcal{N} 
	\left(\frac{E_{\nu}}{\langle E^0_{\nu} \rangle}\right)^{\alpha_{\nu}} \exp\left[-\left(\alpha_{\nu} + 1\right)\frac{E_{\nu}}{\langle E^0_{\nu} \rangle}\right] \ ,
\end{equation}
where $\alpha_{\nu}$ characterizes the pinching and \[\mathcal{N} = \frac{(\alpha_{\nu}+1)^{\alpha_{\nu}+1}}{\langle E_{\nu}\rangle\Gamma(\alpha_{\nu}+1)}\] is the normalization constant, in which $\Gamma(p)$ is 
the Euler Gamma 
function.\footnote{We call the parameters $\eta_{\nu}$ and $\alpha_{\nu}$ simply as 
pinching parameters since their variation directly either suppresses or enhances the 
high energy tails of the distributions.}

In the future, it  would be very valuable to extract, from 
supernova simulations, more information on the pinching parameter (and on a possible non-thermal component) of the neutrino energy distributions. 
It is indeed one of the goals of the present work to emphasize the importance of the
tail in determining the response of detectors having neutrino detection channels with high energy thresholds such as
the two-neutron one in HALO or neutrino-carbon scattering in a scintillator detector like LENA \cite{Wurm:2011zn}. 
Obviously, if one sticks to the above-mentioned parametrizations (\ref{eq:F0}-\ref{eq:PLdistr}), as we do in the following, the primary supernova neutrino fluxes are fully
determined if the neutrino luminosity, the average energy and the pinching parameter are identified. 

\section{Numerical results}
The goal here is to present predictions for the charged-current and neutral current events in a lead based detector like the Helium And Lead Observatory (HALO). The detector is currently under construction and should start to be operating soon~\cite{HALOweb}. In phase-I HALO is made of 79 tons of lead while in phase-II the mass is planned to be increased to 1 kton. Such a detector exploits the following detection channels:  
\begin{equation}
\begin{split}
 \nu_e + {}^{208}\mathrm{Pb} &\rightarrow {}^{207}\mathrm{Bi} + n + e^- \\
 \nu_e + {}^{208}\mathrm{Pb} &\rightarrow {}^{206}\mathrm{Bi} + 2 n + e^- \\
 \nu_x + {}^{208}\mathrm{Pb} &\rightarrow {}^{207}\mathrm{Pb} + n + \nu_x \\
 \nu_x + {}^{208}\mathrm{Pb} &\rightarrow {}^{206}\mathrm{Pb} + 2 n + \nu_x
\end{split}
\end{equation}
where the neutrons are detected using ${}^{3}\mathrm{He}$ counters as done for the SNO experiment \cite{Aharmim:2008kc}.
HALO will measure neutral- and charged-current events without the capability of distinguishing them since the outgoing electrons are not detected. In phase-I the one-neutron ($1n$-) and two-neutron ($2n$-) detection efficiencies are about 50 \% and 25 \% (for seeing both neutrons\footnote{If both neutrons are captured efficiency is nearly 100 \% due to their small separation in time. Additionally, in practice, some $2n$-events can give $1n$-signal if the other neutron escapes the detector. Moreover, secondary neutron production can be seen as a $2n$-signal, although unlikely. Such corrections should be determined by measuring the detector response using HALO phase-I.}), respectively, while the detector will have a good time resolution of about 30 ms~\cite{Virtue}. In the case of a supernova explosion, HALO will have the capability of following the different phases of an explosion, from accretion phase to the cooling of the proto-neutron star. The specificity of measuring $1n$- and $2n$-events gives to such a detector an interesting physics potential, as we will discuss in the following. 

\subsection{Choice of the parameters}
\label{sec:choice}
We consider that a core-collapse supernova explosion occurs in our galaxy, at about 10 kpc from the Earth. For the neutrino fluxes at the neutrinospheres $F^0$ in eq.~\eqref{e:SNF}, both Fermi-Dirac eqs.~\eqref{eq:FDdistr}-\eqref{eq:Fint} and a power-law spectra eq.~\eqref{eq:PLdistr} have been investigated. As an example, we consider luminosity values corresponding to later cooling phase of the proto-neutron star. In order to test whether the energy equipartition is satisfied or not during this time period, we consider here two possibilities: an equal luminosity case with $L\ {}_{\accentset{\left(-\right)}{\nu}_{\l}} = 1/6 \times 10^{52}\ \mathrm{erg\ s}^{-1}$ 
and an extreme case of $L_{\nu_{x}} = 2 L_{\nu_e}$ with $L_{\nu_e} = L_{\bar{\nu}_e} = 1 \times 10^{51}\ \mathrm{erg\ s}^{-1}$, such that the total neutrino luminosity $L^{\rm TOT}_{\nu} = L_{\nu_e} + L_{\bar{\nu}_e} + 4 L_{\nu_{x}} = 10^{52}\ \mathrm{erg\ s}^{-1}$ is the same in both cases in order to be able to compare them.\footnote{As suggested by figure 1 in~ref.~\cite{Choubey:2010up}, when $0.5 \lesssim L_{\nu_x}/L_{\nu_e} \lesssim 1$ there is no qualitative difference in neutrino spectra in collective region within the considered range of neutrino average energies. Therefore, the other extreme case of $L_{\nu_{x}} = 0.5 L_{\nu_e}$ is qualitatively the same as our equal luminosity case. Although, quantitatively our results in numerical section do depend on the exact values of luminosities.} Here(after) we refer to the all non-electron type neutrinos simply by $\nu_x$ (unless otherwise mentioned).

The primary neutrino average energies 
satisfy the hierarchy ${\langle E^0_{\nu_e} \rangle} \lesssim {\langle E^0_{\bar{\nu}_e} \rangle} \lesssim {\langle E^0_{\nu_{x}} \rangle}$, as expected from the difference among the electron (anti)neutrino and the muon and tau (anti)neutrino interaction cross sections with matter, and also from the fact that the medium is neutron-rich. In our calculations we have fixed $\langle E^0_{\nu_e} \rangle = 10\ \mathrm{MeV}$, $\langle E^0_{\bar{\nu}_e} \rangle = 13\ \mathrm{MeV}$ and for the non-electron-type neutrinos we consider the values $\langle E^0_{\nu_{x}} \rangle = 13, 15, 18, 21$ and $25\ \mathrm{MeV}$. Indeed although fits to the SN1987A data appear to show rather low energy components of the fluxes, we wish to explore the sensitivity of our findings to the variation of such input parameters, without sticking to any particular model, having in mind that, e.g. stiffer equations of state for the neutron star might give higher average energies \cite{Sumiyoshi:2007pp}. 

\begin{figure}
\includegraphics[width=1.\textwidth]{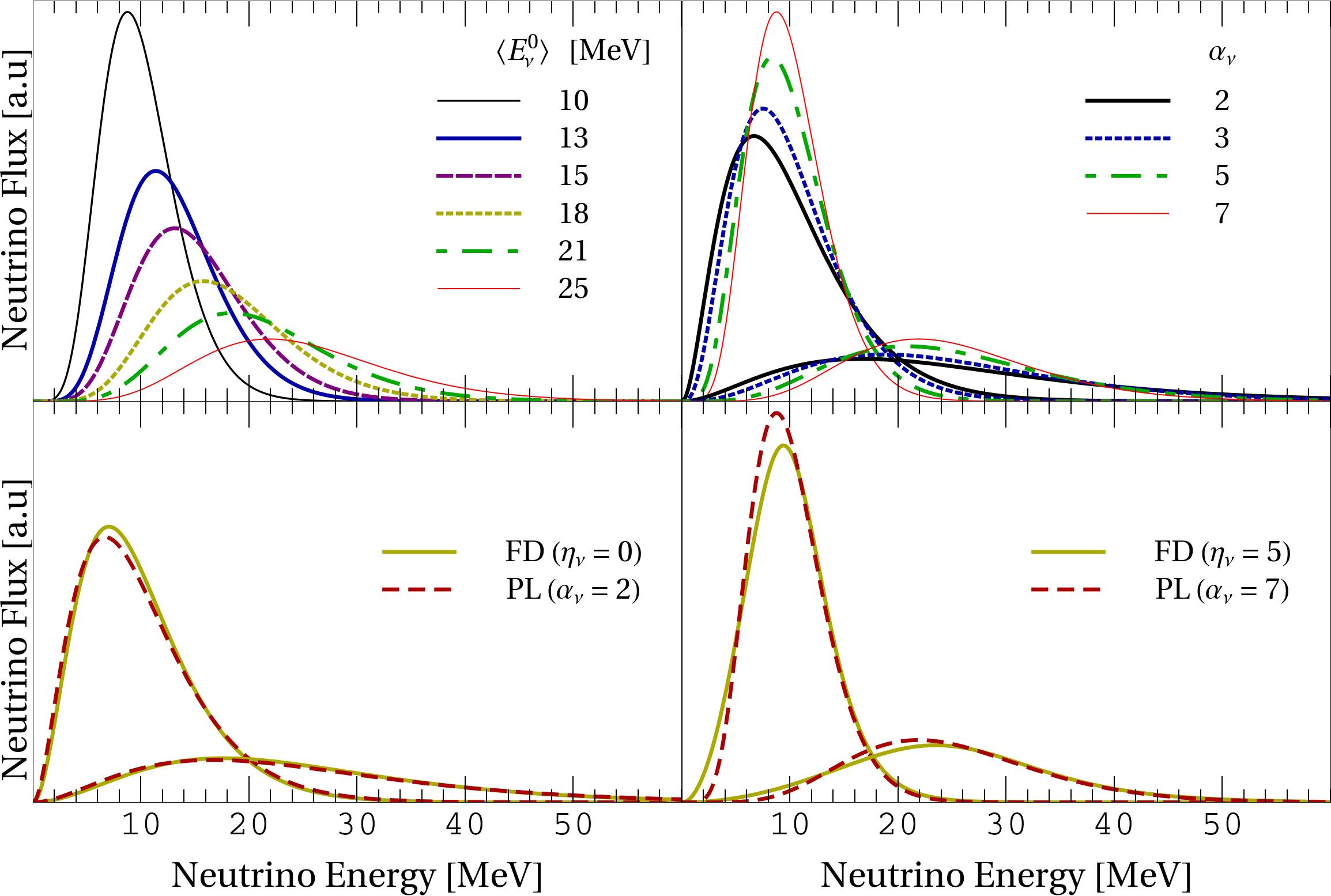}
 \caption{Neutrino energy spectrum at the neutrinosphere. Top left: Power-law (PL) spectrum, eq.~\eqref{eq:PLdistr}, with $\alpha_{\nu}$ fixed to value 7 and average energies, $\langle E^0_{\nu} \rangle$, vary from $10$ to $25\ \mathrm{MeV}$. Top right: PL spectrum with $\alpha_{\nu}$ values varying from $2$ to $7$ for two different average energies ($10$ and $25\ \mathrm{MeV}$). 
Lower figures: Comparison of the initial neutrino Fermi-Dirac (FD), eq.~\eqref{eq:FDdistr} (solid), and power-law (PL, dashed lines) energy spectra for two different average energies ($10$ and $25\ \mathrm{MeV}$). Bottom Left: $\eta_{\nu} = 0$ for FD and $\alpha_{\nu} = 2$ for PL. Bottom right: $\eta_{\nu} = 5$ and $\alpha_{\nu} = 7$.}
  \label{fig:FiniPL_a2and7_Eav_10to25}
\end{figure}

In all of the cases considered we have fixed the electron-type parameters 
to $\alpha_{\nu_e, \bar{\nu}_e} = 3$ and $\eta_{\nu_e, \bar{\nu}_e} = 1$ 
and vary the non-electron-type parameters model independently: 
$\eta_{\nu_x} \in [0,5]$ and $\alpha_{\nu_x} \in [2,7]$. 
Note that in ref.~\cite{Keil:2002in} all the neutrino fluxes are found to be characterized by pinching parameters with values $\alpha_{\nu} \in [2,5]$ for power-law, eq.~\eqref{eq:PLdistr}, or $\eta_{\nu} \in [0,4]$ for Fermi-Dirac distribution, eq.~\eqref{eq:FDdistr}.
We have tested that varying the electron-type pinching parameter values has only a very small or no effect in all of the cases we consider. Currently still little information on such parameters can be found in the literature from supernova simulations. However it is clear that to properly define neutrino fluxes one needs the first as well as the second 
moment of the distribution \cite{Keil:2002in,Myra:1990tt}. Figure \ref{fig:FiniPL_a2and7_Eav_10to25} presents the neutrino fluxes at the neutrinosphere for the two parametrizations considered. Since the results we obtain are qualitatively and quantitatively very similar in the following we show results obtained for the power-law spectra only. 

As far as the neutrino masses and mixing are concerned, most of such parameters are now known, in particular $\Delta m^2_{21} = (7.59 \pm 0.20) \times 10^{-5}\ \mathrm{eV}^2$, $|\Delta m^2_{32}| = (2.43 \pm 0.13) \times 10^{-3}\ \mathrm{eV}^2 (\approx |\Delta m^2_{31}|)$, $\sin^{2} 2\theta_{12} = 0.87 \pm 0.03$, $\sin^{2} 2\theta_{23} > 0.92$ and $\sin^{2} 2\theta_{13} < 0.15$~\cite{Nakamura:2010zzi}. For the mixing angles we have used the values $\sin^{2} 2\theta_{12} = 0.86$, $\sin^{2} 2\theta_{23} = 0.99$ while for the third angle our results are obtained either considering that $\theta_{13}$ is very small $(0.001)$ 
or close to the present Chooz limit ($0.1$). We shall consider both normal (inverse) hierarchy coming from a positive (negative) sign for $|\Delta m^2_{23}|$. 

Note that the measured values of the solar parameters make the flavor transition at the low resonance always adiabatic. That is, in eq.~\eqref{eq:PMSW} $P_L = 0$ for typical density profiles from supernova simulations, while 
the adiabaticity of the high resonance crossing depends on the still unknown third neutrino mixing angle $\theta_{13}$. 
Our choices for the $\theta_{13}$ imply that all the processes are assumed to be completely adiabatic $P_H = 0$ (large) or non-adiabatic $P_H = 1$ (very small $\theta_{13}$). We set the Dirac CP phase to zero. For a discussion of its impact see \cite{Balantekin:2007es,Gava:2008rp}.

\subsection{Computation of neutrino fluxes}
For the $P_{\nu\nu}$ matrix eq.~\eqref{eq:Pnunu} we parametrize the neutrino fluxes after the region where the neutrino-neutrino interactions are important following the results of the full three flavor numerical calculations with/without equipartition for the neutrino luminosity~\cite{Choubey:2010up,Dasgupta:2010cd,Mirizzi:2010uz}. These references show that after the collective effects there can be no splits, one low (or high) energy split or both low and high energy splits in (anti)neutrino spectra, depending on the neutrino flux parameters and mass hierarchy.\footnote{The current status for the appearance of the splits in neutrino energy spectrum and their dependence on neutrino luminosities and mass hierarchy can be found e.g. in figure 1 of ref.~\cite{Choubey:2010up}.} In our considerations this implies that in eq.~\eqref{eq:Pnunu} we take
\begin{subequations}
	\label{eqs:P1P2P3}
	\begin{align}
		\label{eq:Pey}
		&P_{ey} =
			\left\{
				\begin{array}{cl}
					1, & \text{if} \ E^s_{l} < E < E^s_{h} \\
					0, & \text{otherwise} \ ,
				\end{array}
			\right. \\
		\label{eq:Pex}
		&P_{ex} = 
			\left\{
				\begin{array}{cl}
					1, & \text{if} \ E > E^s_{h} \\
					0, & \text{otherwise} \ ,
				\end{array}
			\right.	
	\end{align}
\end{subequations}
where, for simplicity, we have assumed all the splits in energy to be sharp i.e. the swapping probabilities are step functions. $E^s_{l}$ and $E^s_{h}$ are the low and high split energies, respectively. 

The above parametrization can be easily modified to correspond other possible cases. For instance, it is to be noticed that the recent multi-angle simulations~\cite{Mirizzi:2010uz} show differences in neutrino spectra after the collective effects compared to the single-angle simulations given in refs.~\cite{Choubey:2010up,Dasgupta:2010cd}. General feature of multi-angle effects is to make splits less sharp in energy. In IMH ref.~\cite{Mirizzi:2010uz} shows that considering neutrino spectra with single split (our equal luminosity case) multi-angle effects are marginal, but for multiple splits
(our $L_{\nu_x} = 2 L_{\nu_e}$ example case) multi-angle effects can be relevant. Namely in IMH with multiple splits one finds strong suppression of $(e-x)$-swapping. In our case for IMH with $L_{\nu_x} = 2 L_{\nu_e}$ this multi-angle result is obtained by setting in eq.~\eqref{eq:Pex} $P_{ex} = 0$ at all energies, that is, there exists only $(e-y)$-swap at intermediate energies and at low and high energies spectra remain equal to the primary.

Using eqs.~\eqref{eqs:P1P2P3} and~\eqref{eq:Pnunu} one can compute the fluxes after the collective region $F' = (F'_{\nu_e}, F'_{\nu_x}, F'_{\nu_y})^T$: 
\begin{equation}
\label{eq:F'}
 F' = P_{\nu\nu} F^0 \ .
\end{equation}
The fluxes are listed in table \ref{tab:Fnunu} in the cases of equal and non-equal ($L_{\nu_x} = 2 L_{\nu_e}$) luminosity both in normal and inverted mass hierarchy. According to~\cite{Choubey:2010up,Dasgupta:2010cd}, in the computation of event numbers we have used the numerical values $E^s_{l} = 8\ \mathrm{MeV}$ and $E^s_{h} = 23\ \mathrm{MeV}$, when appropriate (see table \ref{tab:Fnunu}).\footnote{For antineutrinos one can use the same form as above by replacing $F^0$ by $\overline{F}^0 = (F^0_{\bar{\nu}_e}, F^0_{\bar{\nu}_{x}}, F^0_{\bar{\nu}_{y}})^T$ and $P_{\nu\nu}$ with $\overline{P}_{\nu\nu}$ such that $P_{\l\l,ex,ey} = \overline{P}_{\l\l,ex,ey}$ and $E^s_{l,h} = \overline{E^s}_{l,h}$.}

\begin{table}[t]
\begin{center}
 \begin{tabular}{l@{\extracolsep{6pt}}cc|cccc}
  \hline
  \multicolumn{6}{c}{$(F'_{\nu_e}, F'_{\nu_x}, F'_{\nu_y})$} \\
  \hline
  & \multicolumn{2}{c|}{Equal Luminosity} & \multicolumn{3}{c}{$L_{\nu_x} = 2 L_{\nu_e}$} \\
  \hline
  \multirow{2}{*}{$E_{\nu}$ range} & \multirow{2}{*}{IMH} & \multirow{2}{*}{NMH} & \multicolumn{2}{c}{IMH} & \multirow{2}{*}{NMH}  \\
  & & & single-angle & multi-angle &  \\ 
  \hline
  $0 - E^s_l$ & $(F^0_{\nu_e}, F^0_{\nu_x}, F^0_{\nu_y})$ & $(F^0_{\nu_e}, F^0_{\nu_x}, F^0_{\nu_y})$ & $(F^0_{\nu_e}, F^0_{\nu_x}, F^0_{\nu_y})$ & $(F^0_{\nu_e}, F^0_{\nu_x}, F^0_{\nu_y})$ & $(F^0_{\nu_e}, F^0_{\nu_x}, F^0_{\nu_y})$  \\
  $E^s_l - E^s_h$ & $(F^0_{\nu_y} , F^0_{\nu_x} , F^0_{\nu_e})$ & $\|$ & $(F^0_{\nu_y}, F^0_{\nu_x}, F^0_{\nu_e})$ & $(F^0_{\nu_y}, F^0_{\nu_x}, F^0_{\nu_e})$ & $\|$ \\
  $E^s_h - \infty$ & $\|$ & $\|$ & $(F^0_{\nu_x}, F^0_{\nu_e}, F^0_{\nu_y})$ & $(F^0_{\nu_e}, F^0_{\nu_x}, F^0_{\nu_y})$ & $(F^0_{\nu_y}, F^0_{\nu_x}, F^0_{\nu_e})$ \\
  %
  \hline
 \end{tabular}
\end{center}
\caption{Neutrino fluxes after the collective region $(F'_{\nu_e}, F'_{\nu_x}, F'_{\nu_y})$ given by eq.~\eqref{eq:F'}, depending on their energy. The fluxes are given in terms of the neutrino fluxes at the neutrinospheres $(F^0_{\nu_e}, F^0_{\nu_x}, F^0_{\nu_y})$, eq.~\eqref{eq:F0}, in equal and non-equal luminosity case, both in normal (NMH) and inverted mass hierarchy (IMH). $E^s_l$ and $E^s_h$ are the values of the low and high split energies, respectively (see text). The $\|$ sign means that the result is the same as above.}
\label{tab:Fnunu}
\end{table}

The exact value of the split energy depends on the neutrino mixing and the primary neutrino flux parameter values. Considering only two neutrino flavors the equation of motion for the system implies the ``conservation of flavor lepton number''(see e.g. ref.~\cite{Fogli:2007bk}):
\begin{equation}
 \label{eq:cfln}
 N_{\nu_e} - N_{\bar{\nu}_e} = D(r) \int \d E_{\nu} \left( F^0_{\nu_e} - F^0_{\bar{\nu}_e} \right) = \text{const} \ ,
\end{equation}
where $N_{\nu_e}(N_{\bar{\nu}_e})$ is the effective number density of $\nu_e(\bar{\nu}_e)$, $D(r)$ includes (flavor independent) geometrical dependence and $F^0_{\nu_{e}, \bar{\nu}_e}$ is given in eq.~\eqref{eq:F0}.

In core-collapse supernovae initial $\nu_e$ number density is larger than that of $\bar{\nu}_e$. Therefore, if all the $\bar{\nu}_e$'s convert to non-electron antineutrinos the conservation of flavor lepton number implies that not all of the $\nu_e$'s can convert to non-electron neutrinos. This condition is met e.g. in the case of equipartition of energy in IMH. Assuming complete $\bar{\nu}_e \leftrightarrow \bar{\nu}_y$ swap in the antineutrino sector, conservation of flavor lepton number in eq.~\eqref{eq:cfln} indicates a single split energy $E^s$ for neutrinos:
\begin{equation}
 \label{eq:Esplit}
 \begin{split}
  \int_{0}^{\infty} \d E_{\nu} \left[ (F^0_{\nu_e} - F^0_{\nu_x}) - (F^0_{\bar{\nu}_e} - F^0_{\bar{\nu}_x}) \right] = &\int_{0}^{E^s} \d E_{\nu} (F^0_{\nu_e} - F^0_{\nu_x}) - \int_{E^s}^{\infty} \d E_{\nu} (F^0_{\nu_e} - F^0_{\nu_x}) \\
  &+ \int_{0}^{\infty} \d E_{\nu} (F^0_{\bar{\nu}_e} - F^0_{\bar{\nu}_x}) \ ,
 \end{split}
\end{equation}
where l.h.s. corresponds to initial and r.h.s. to final states, giving
\begin{equation}
 \label{eq:Esplit2}
 \int_{E^s}^{\infty} \d E_{\nu} (F^0_{\nu_e} - F^0_{\nu_x}) = \int_{0}^{\infty} \d E_{\nu} (F^0_{\bar{\nu}_e} - F^0_{\bar{\nu}_x}) \ .
\end{equation}

We have tested that using eq.~\eqref{eq:Esplit2} the value of $E^s$ varies from $6$ 
to $10\ \mathrm{MeV}$ when changing the primary neutrino flux parameter values 
($\eta$ or $\alpha_{\nu}$, $\langle E^0_{\nu} \rangle$) within the ranges given 
in the previous subsection. Such a modification of $E^s$ has no consequence on the results presented because of the location of the $1n$ energy threshold.
Therefore, we have kept the split energy values 
fixed in all the cases. On the other hand, we have checked that
considering a similar variation of $\pm 2\ \mathrm{MeV}$ 
for the higher energy split has only a small effect on the numbers of events in a ${\rm Pb}$ detector that we will show. 

The (primed) fluxes after the collective region in eq.~\eqref{eq:F'} enter the MSW region. The mass eigenstate fluxes $F_{\nu_i}$ 
which exit the surface, can also be calculated 
using eqs.~\eqref{e:SNF} and~\eqref{eq:PMSW}:
\begin{equation}
 \label{eq:FMSW}
  	\begin{pmatrix}
		F_{\nu_1} \\
		F_{\nu_2} \\
		F_{\nu_3}
	\end{pmatrix} = P_{MSW}
	\begin{pmatrix}
		F'_{\nu_e} \\
		F'_{\nu_x} \\
		F'_{\nu_y}
	\end{pmatrix} \ .
\end{equation}
As pointed out earlier, the fate of the neutrino fluxes in the MSW region depends on the neutrino mass hierarchy and the adiabaticity of the resonance crossings. For neutrinos in normal mass hierarchy for small $\theta_{13}\ (P_H = 1)$ and large $\theta_{13}\ (P_H = 0)$ the neutrino mass fluxes which exit the star are given by
\begin{equation}
	\label{eq:MSWnmhSMALL}
	\begin{pmatrix}
		F_{\nu_1} \\
		F_{\nu_2} \\
		F_{\nu_3}
	\end{pmatrix} =
	\left\{
		\begin{array}{cl}
			\begin{pmatrix}
				F'_{\nu_x} \\
				F'_{\nu_e} \\
				F'_{\nu_y}
			\end{pmatrix} \quad \text{for small }\theta_{13} \ . \\
			\begin{pmatrix}
				F'_{\nu_x} \\
				F'_{\nu_y} \\
				F'_{\nu_e}
			\end{pmatrix} \quad \text{for large }\theta_{13} \ .
		\end{array}
	\right.
\end{equation}
In IMH $P_H = 1$ regardless of the value of $\theta_{13}$. Therefore, in the absence of collective effects the case of inverted is degenerate with normal mass hierarchy and small $\theta_{13}$.

\begin{figure}
 \includegraphics[width=1.0\textwidth]{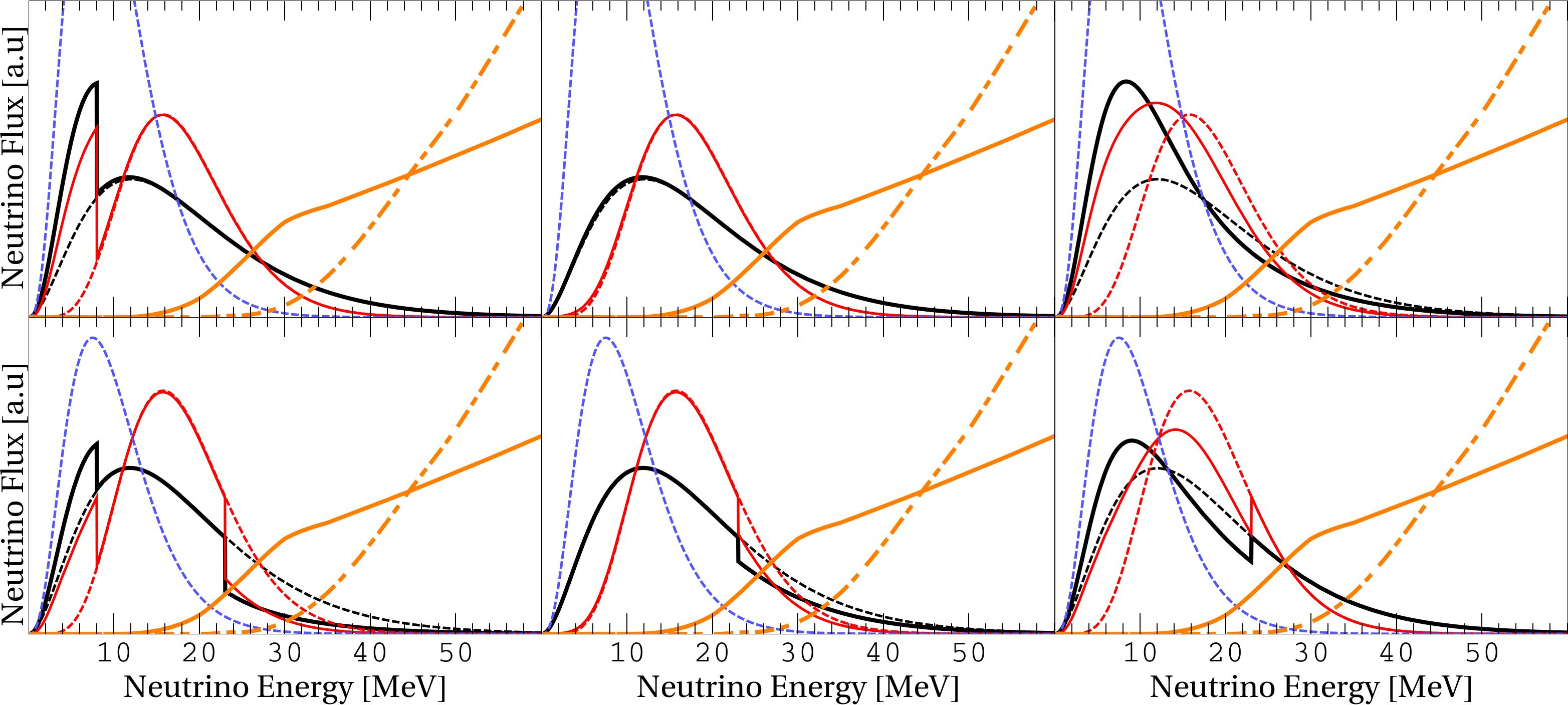}
\caption{(Color online) Electron neutrino fluxes at Earth, eq.~\eqref{eq:Fe}, (solid lines) as a function of energy including $\nu-\nu$ interactions, the MSW effect and decoherence. The primary $\nu_e$ (blue thin dashed) and $\nu_{x}$ (black and red thin dashed) fluxes at the neutrinosphere, eqs.~\eqref{eq:F0} and \eqref{eq:PLdistr}, are also shown. Black (red) lines correspond to pinching parameter $\alpha_{\nu_{x}}$ = 2 (7). For the primary $\nu_e$ flux $\alpha_{\nu_{e}}$ = 3. In this figure the primary average energies are fixed as $\langle E^0_{\nu_e} \rangle = 10\ \mathrm{MeV}, \langle E^0_{\nu_{x}} \rangle = 18\ \mathrm{MeV}$. Upper row: equal luminosities, lower row: $L_{\nu_x} = 2 L_{\nu_e}$. Left panel: inverted mass hierarchy (IMH), middle panel: normal mass hierarchy (NMH) with large $\theta_{13}$, right panel: NMH with small $\theta_{13}$. Additionally, the charge current $\nu_e - \mathrm{Pb}$ one-(thick, solid, orange) and two-neutron (thick, dash-dotted, orange) emission cross sections (from ref.~\cite{Engel:2002hg}) are also shown.}
 \label{fig:Fe}
\end{figure}

The observable neutrino fluxes at the Earth are now obtained from the mass eigenstate fluxes by using the matrix $A$ in eq.~\eqref{eq:A}. 
That is, for instance, the final electron neutrino flux at the Earth is given by
\begin{equation}
 \label{eq:Fe}
 F_{\nu_{e}} = \sum_{i = 1}^{3} \left| U_{e i} \right|^2 F_{\nu_i} \ ,	
\end{equation}
where using $\theta_{13} = 0.001(0.1)$
\begin{equation}
\label{eq:Uelem}
	\begin{split}
		&\left| U_{e 1} \right|^2 \approx 0.69(0.68) \ , \\ 
		&\left| U_{e 2} \right|^2 \approx 0.31(0.31) \ , \\ 
		&\left| U_{e 3} \right|^2 \approx 10^{-6}(0.01) \ ,
	\end{split}
\end{equation}
giving the weights of the mass eigenstate fluxes which exit the surface. Table~\ref{tab:Ffinal} shows the energy dependence of the mass eigenstate fluxes in terms of the primary neutrino fluxes in all the scenarios considered. In figure~\ref{fig:Fe} the corresponding final electron neutrino fluxes (using the power law distribution in eq.~\eqref{eq:PLdistr} for the primary fluxes) are plotted as a function of energy, also showing the dependence of the spectral shape on the value of $\alpha_{\nu_x}$. 
The figure also presents the neutron emission charged-current $\nu_e - \mathrm{Pb}$ cross sections, to show which part of the $\nu_e$ spectrum detected on Earth is relevant from the point of view of neutrino detection on a lead observatory.

\begin{table}[t]
	\begin{center}
		\begin{tabular}{l@{\extracolsep{6pt}}cccc}
			\hline
			\multicolumn{4}{c}{$(F_{\nu_1},\ F_{\nu_2},\ F_{\nu_3})$} \\
			\hline
			\multirow{2}{*}{$E_{\nu}$ range} & \multirow{2}{*}{IMH} & \multicolumn{2}{c}{NMH} \\ 
			& & large $\theta_{13}$ & small $\theta_{13}$  \\
			\hline
			$0 - E^s_l$ & $(F^0_{\nu_x}, F^0_{\nu_e}, F^0_{\nu_y})$ & $(F^0_{\nu_x}, F^0_{\nu_y}, F^0_{\nu_e})$ & $(F^0_{\nu_x}, F^0_{\nu_e}, F^0_{\nu_y})$ \\ 
			$E^s_l - E^s_h$ & $(F^0_{\nu_x}, F^0_{\nu_y}, F^0_{\nu_e})$ & $\|$ & $\|$ \\
			$E^s_h - \infty$ & $\|$ & $\|$ & $\|$ \\
			\hline
		\end{tabular}
	\end{center}
\caption{Neutrino mass eigenstate fluxes $(F_{\nu_1}, F_{\nu_2}, F_{\nu_3})$, eq.~\eqref{eq:FMSW}, depending on neutrino energy, exiting the star after the MSW resonance region in equal luminosity case.  The fluxes are given in terms of the neutrino fluxes at the neutrinospheres $(F^0_{\nu_e}, F^0_{\nu_x}, F^0_{\nu_y})$, eq.~\eqref{eq:F0}, both in normal (NMH) and inverted mass hierarchy (IMH). 
Notice that in normal mass hierarchy these results depend on the value of the third neutrino mixing angle $\theta_{13}$. 
$E^s_l$ and $E^s_h$ are the values of the low and high split energies, respectively (see text). The $\|$ sign means that the result is the same as above.
	}
	\label{tab:Ffinal}
\end{table}

\subsection{Computation of event rates}
\noindent
Since the detector we will consider does not identify the outgoing lepton, to compute the total event rate $N^{\rm TOT}$ we have to add  charged-current ($CC$) and
neutral-current ($NC$) event rates 
\begin{equation}
 \label{eq:Ntot}
 N^{\rm TOT}_{1n(2n)} = N^{CC}_{1n(2n)} + N^{NC}_{1n(2n)} \ ,
\end{equation} 
where $N_{1n(2n)}$ refers to 1$n$(2$n$) event rates. The CC and NC event rates are given by
\begin{eqnarray}
 \label{eq:NCC}
 N^{CC}_{1n(2n)} &=& N_{\rm Pb} \int \d E {\frac{\d \sigma^{CC}_{1n(2n)}}{\d E}} F_{\nu_e}(E) \ , \\
 \label{eq:NNC}
 N^{NC}_{1n(2n)} &=& N_{\rm Pb}\int \d E \sum_{\l=e,\mu,\tau} \nonumber \\
 & & \times \left[ {\frac{\d \sigma^{NC,\nu}_{1n(2n)}}{\d E}}  F_{\nu_{\l}}(E) + {\frac{\d \sigma^{NC,\bar{\nu}}_{1n(2n)}}{\d E}} F_{\bar{\nu}_{\l}}(E) \right] , 
\end{eqnarray}
with $N_{\rm Pb}$ the number of target particles in lead detector\footnote{Here we consider 1 kton lead made of $^{208}$Pb so that $N_{\rm Pb}= 2.9\times10^{30}$. Our results can be easily scaled to natural composition.}, 
$\d \sigma_{1n(2n)}/\d E$ the differential cross sections and $F\ {}_{\accentset{\left(-\right)}{\nu}_{\l}}$ the final (anti)neutrino flux given by eq.~\eqref{e:SNF}.\footnote{Notice that,
$\sum_{\l = e,\mu,\tau} F\ {}_{\accentset{\left(-\right)}{\nu}_{\l}} = \sum_{\l = e,\mu,\tau} F^0\ {}_{\accentset{\left(-\right)}{\nu}_{\l}}$, that is the NC events are not affected by the flavor transformations.}

As an example, we consider the case of inverted mass hierarchy (single-angle) with $L_{\nu_x} = 2 L_{\nu_e}$. According to eqs.~\eqref{eq:Fe} and \eqref{eq:NCC} (results listed in table~\ref{tab:FfinalneqL}) the charged-current event rates are explicitly calculated in terms of primary neutrino fluxes given in eq.~\eqref{eq:F0} as follows:
\begin{equation}
 \label{eq:NCCex}
 \begin{split}
  \frac{1}{N_{\rm Pb}} N^{CC}_{1n(2n)} = &\int_0^{E^s_{l}} \d E {\frac{\d \sigma^{CC}_{1n(2n)}}{\d E}} \left( \left| U_{e 1} \right|^2 F^0_{\nu_x} + \left| U_{e 2} \right|^2 F^0_{\nu_e} + \left| U_{e 3} \right|^2 F^0_{\nu_y} \right) \\
  &+ \int_{E^s_{l}}^{E^s_{h}} \d E {\frac{\d \sigma^{CC}_{1n(2n)}}{\d E}} \left( \left| U_{e 1} \right|^2 F^0_{\nu_x} + \left| U_{e 2} \right|^2 F^0_{\nu_y} + \left| U_{e 3} \right|^2 F^0_{\nu_e} \right) \\
  &+ \int_{E^s_{h}}^{\infty} \d E {\frac{\d \sigma^{CC}_{1n(2n)}}{\d E}} \left( \left| U_{e 1} \right|^2 F^0_{\nu_e} + \left| U_{e 2} \right|^2 F^0_{\nu_x} + \left| U_{e 3} \right|^2 F^0_{\nu_y} \right) \ ,
 \end{split}
\end{equation}
where the weights $\left| U_{e i} \right|^2$ are given in eq.~\eqref{eq:Uelem}. Note that, according to our assumptions, $F^0_{\nu_y} = F^0_{\nu_x}$.

\begin{table*}[t]
	
	\begin{center}
		\begin{tabular}{ccccc}
			\hline
			\multicolumn{5}{c}{$(F_{\nu_1},\ F_{\nu_2},\ F_{\nu_3})$} \\
			\hline
			\multirow{2}{*}{$E_{\nu}$ range} & \multicolumn{2}{c}{IMH} & \multicolumn{2}{c}{NMH} \\
			& single-angle & multi-angle & large $\theta_{13}$ & small $\theta_{13}$ \\
			\hline
			$0 - E^s_l$ & $(F^0_{\nu_x}, F^0_{\nu_e}, F^0_{\nu_y})$ & $(F^0_{\nu_x}, F^0_{\nu_e}, F^0_{\nu_y})$ & $(F^0_{\nu_x}, F^0_{\nu_y}, F^0_{\nu_e})$ & $(F^0_{\nu_x}, F^0_{\nu_e}, F^0_{\nu_y})$ \\
			$E^s_l - E^s_h$ & $(F^0_{\nu_x}, F^0_{\nu_y}, F^0_{\nu_e})$ & $(F^0_{\nu_x}, F^0_{\nu_y}, F^0_{\nu_e})$ & $\|$ & $\|$ \\
			$E^s_h - \infty$ & $(F^0_{\nu_e}, F^0_{\nu_x}, F^0_{\nu_y})$ & $(F^0_{\nu_x}, F^0_{\nu_e}, F^0_{\nu_y})$ & $(F^0_{\nu_x}, F^0_{\nu_e}, F^0_{\nu_y})$ & $(F^0_{\nu_x}, F^0_{\nu_y}, F^0_{\nu_e})$ \\
			\hline
		\end{tabular}
	\end{center}
\caption{Neutrino mass eigenstate fluxes $(F_{\nu_1}, F_{\nu_2}, F_{\nu_3})$, eq.~\eqref{eq:FMSW}, depending on neutrino energy, exiting the star after the MSW resonance region but for the $L_{\nu_x} = 2 L_{\nu_e}$ case.  The fluxes are given in terms of the neutrino fluxes at the neutrinospheres $(F^0_{\nu_e}, F^0_{\nu_x}, F^0_{\nu_y})$, eq.~\eqref{eq:F0}, both in normal (NMH) and inverted mass hierarchy (IMH). 
Notice that in normal mass hierarchy these results depend on the value of the third neutrino mixing angle $\theta_{13}$. 
$E^s_l$ and $E^s_h$ are the values of the low and high split energies, respectively (see text). The $\|$ sign means that the result is the same as above. 
	}
	\label{tab:FfinalneqL}
\end{table*}

The neutrino-lead total ($0 + 1n + 2n$), $1n$- and $2n$-emission cross sections we use are from a microscopic calculation based on the Random Phase Approximation (RPA) (see table I of ref.\cite{Engel:2002hg}).\footnote{The electron anti-neutrino detection channel is strongly suppressed due to Pauli blocking.} These calculations include nuclear matrix elements up to multipolarity 4 and assume neutrons emitted with a 100 \% probability once the neutrino 
energy is above the $1n$- and the $2n$-  emission thresholds\footnote{In \cite{Engel:2002hg} it is assumed that the probability for emitting 
one (two) neutron(s) is 100 \% if the neutrino energy is above 10 MeV (20 MeV).}. A refinement of such a calculation would require the inclusion of all possible decay channels with a statistical code, as done in \cite{Kolbe:2000np}. 

\begin{figure}
 \begin{center}
 \includegraphics[width=.7\textwidth]{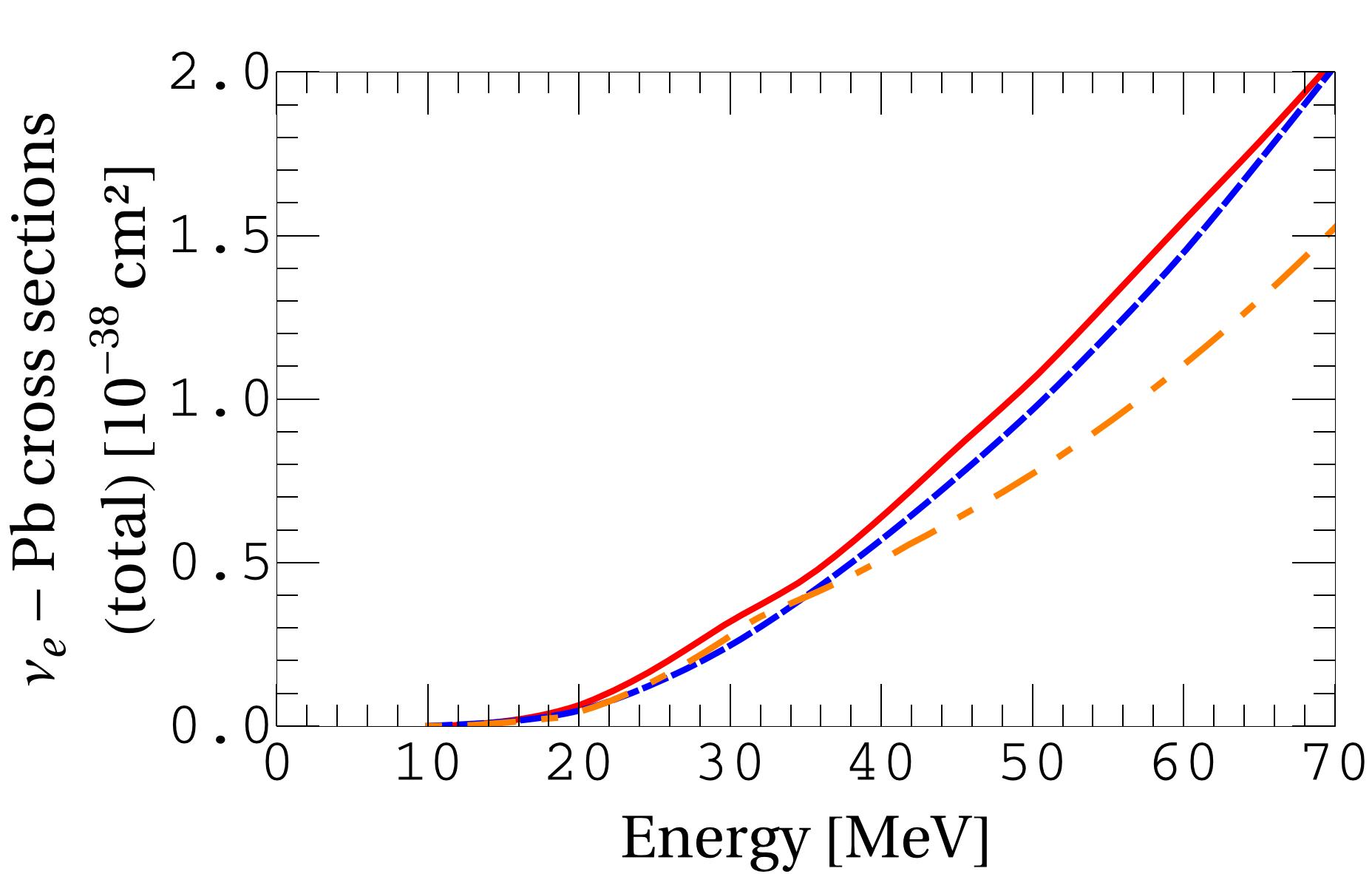}
  \caption{$\nu_e - {\rm Pb}$ charged current total $(0 + 1n + 2n)$ cross sections as a function of energy; RPA~\cite{Engel:2002hg} (solid), CRPA~\cite{Kolbe:2000np} (dashed) and RRPA~\cite{Paar} (dash-dotted).}
  \label{fig:CSs}
 \end{center}
\end{figure}

\begin{figure}
\includegraphics[width=1.\textwidth]{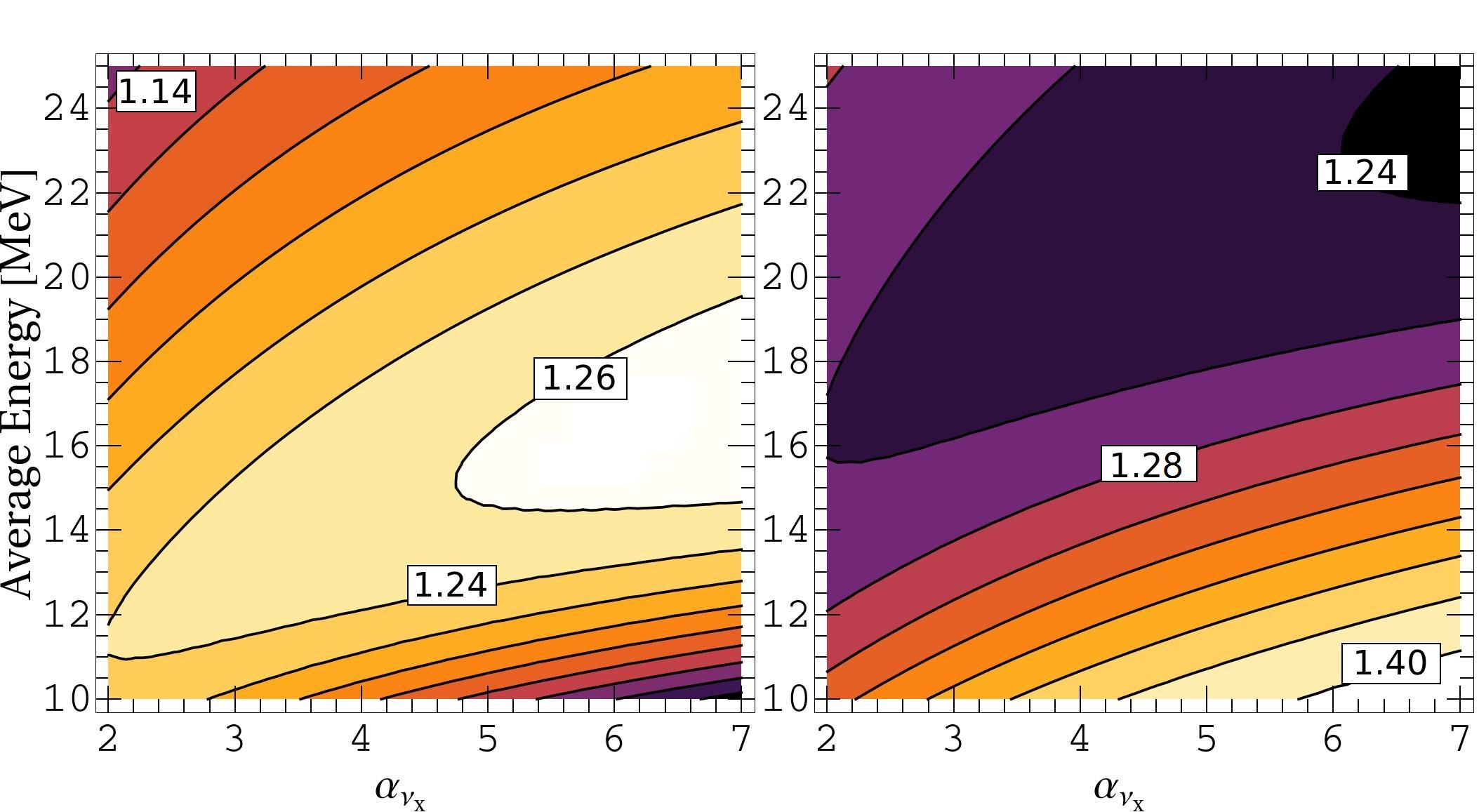}
	\caption{Contour plots of ratios of flux averaged total (0 + 1$n$ + 2$n$) cross sections. Left: RPA~\cite{Engel:2002hg} to CRPA~\cite{Kolbe:2000np}. Contours correspond to ratios from 1.1 to 1.26 (outside to inside) with step of 0.02. Right: RPA to RRPA~\cite{Paar}, similarly from 1.4 to 1.24 with step of 0.02.}
	\label{fig:AveAlphaContours}
\end{figure}

In order to show the current uncertainty on the neutrino-lead cross sections figure~\ref{fig:CSs} presents a comparison of the total (0 + $1n$ + $2n$) neutron emission cross sections with those of refs.~\cite{Kolbe:2000np, Paar} based on Continuum RPA (CRPA) and Relativistic RPA (RRPA), respectively. Figure~\ref{fig:AveAlphaContours} shows ratios of flux-averaged neutrino-lead cross sections with the neutrino fluxes given by the power-law distribution with varying average energies and pinching parameters. One can see that, even for low average energies, the uncertainty coming from variances of the current theoretical predictions ranges from about 1.2 to 1.4. In the following we will show that, still interesting information can be extracted with a lead-based detector, in spite of the existent cross section uncertainties.

In all of our calculations we assume a 100 \% detection efficiency, so that our results can be scaled easily, depending on the actual neutron detection efficiencies\footnote{Strictly speaking, detection efficiencies do have some dependence on emitted neutron energy, although it is likely to be small. A refinement of the present calculations would require to determine the neutron energy spectrum as done in \cite{Kolbe:2000np}}. The results shown in the next section correspond to HALO phase-II $(1\ \mathrm{kton})$, while those for HALO phase-I can be obtained by multiplying the rates by a factor of $0.079$.\footnote{Note that, if the absolute value of luminosities, detection efficiency, size of the detector or distance to supernova are different from what considered here, one can just multiply our results with a suitable factor as long as the change does not pose qualitative differences.} 

\subsection{Expected events in HALO}
\label{sec:events}
\noindent
We present new predictions for the $1n$- and $2n$-events going beyond the previous calculations \cite{Fuller:1998kb,Engel:2002hg} and include collective effects, 
assuming an explosion in our galaxy. In particular, the results shown are : i) for event rates ; ii) for total number of events ; iii) for ratios of events.
Since HALO has a good time resolution, the event rates are given in order to make time dependent predictions that should be multiplied by the time duration of the corresponding phase in the explosion. In such a period the neutrino flux parameters are assumed to be fixed. Then we show, as an example, the expected numbers of events during the 
accretion and cooling phases and from the whole explosion. Finally, ratios of one- and two-neutron events are given since these turn out to be particularly sensitive to the primary neutrino fluxes. Besides they have the advantage that they are free from common normalization parameters, like the total luminosity. 

Our goal here is two-fold. First, we want to give updated predictions for the HALO project under construction at SNOLAB. Second, we investigate which information on the neutrino fluxes at the neutrinosphere and/or on the neutrino properties can be extracted with such an observatory. For example, it is \textit{a priori} not clear that information can be extracted from the primary pinching parameter, after the inclusion of the collective effects and with the current design of the HALO detector. Indeed it was first pointed out in ref.~\cite{Engel:2002hg} that the measurement of the energy of the outgoing lepton would help constraining the primary pinching parameter. As we will discuss, by considering combinations of $1n$- and $2n$-events some information on the primary neutrino fluxes can indeed be obtained. 
 
Before presenting our results we emphasize that, in order to extract as much information as possible from a future measurement, it will be indeed crucial to combine results from the HALO detector with those from detectors based on other technologies, like water Cherenkov (e.g. Super-Kamiokande or MEMPHYS), scintillators (Borexino, LVD, KamLAND, LENA) or liquid argon (GLACIER) \cite{Autiero:2007zj}. These detectors will be sensitive to electron anti-neutrinos through inverse beta decay and electron neutrinos using scattering on nuclei (oxygen, carbon, argon). In particular, the measurement
of neutral current events through scattering on electrons or protons \cite{Beacom:2002hs} can be used to determine the total neutrino luminosity and possibly also extract the spectra of non-electron type neutrinos.
Note also that the number of unknowns considered in the present work might be narrowed down in the near future
if the third neutrino mixing angle is determined by the reactor Double-Chooz, Daya-Bay,
RENO or T2K and NO$\nu$A experiments \cite{Mezzetto:2010zi}.

\begin{figure*}
\includegraphics[width=1.0\textwidth]{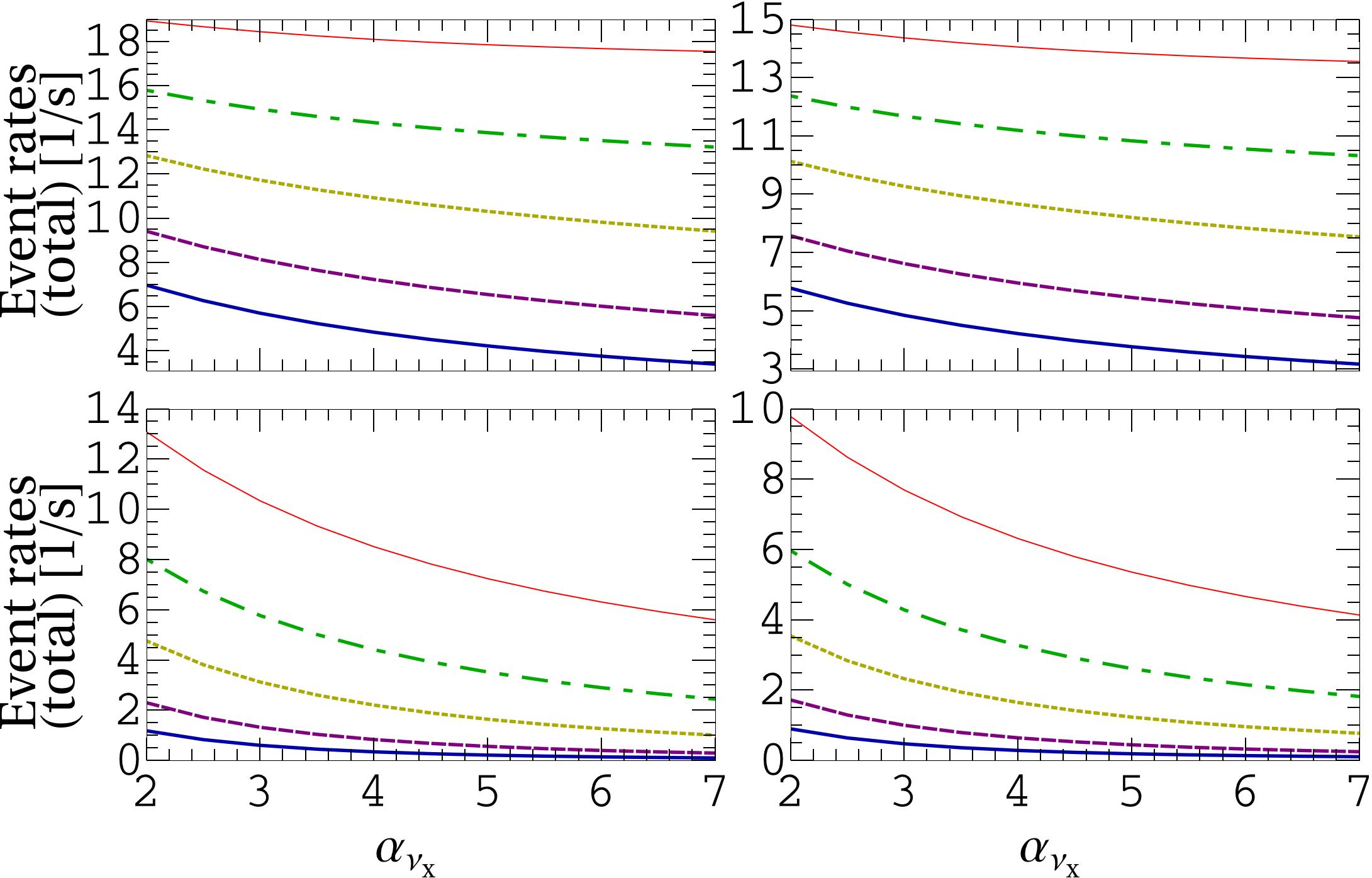}	
	\caption{One- (upper) and two-neutron (lower row) total (charged + neutral current) event rates as a function of primary non-electron neutrino pinching parameter $\alpha_{\nu_x}$, including $\nu-\nu$ interactions, the MSW effect and decoherence. Here is assumed 100 \% detection efficiency and equal luminosities, $L_{\nu}^{\mathrm{TOT}} = 10^{52}\ \mathrm{erg\ s^{-1}}$, in IMH (left) and NMH with small $\theta_{13}$ (right panel). Distance to the supernova is taken to be 10 kpc and target mass 1 kton of ${}^{208}$Pb. Different lines correspond to different non-electron-type primary neutrino average energies (from bottom to top: $13, 15, 18, 21$ and $25\ \mathrm{MeV}$).}
	\label{fig:resALLeqL}
\end{figure*}

\begin{figure*}
\includegraphics[width=1.0\textwidth]{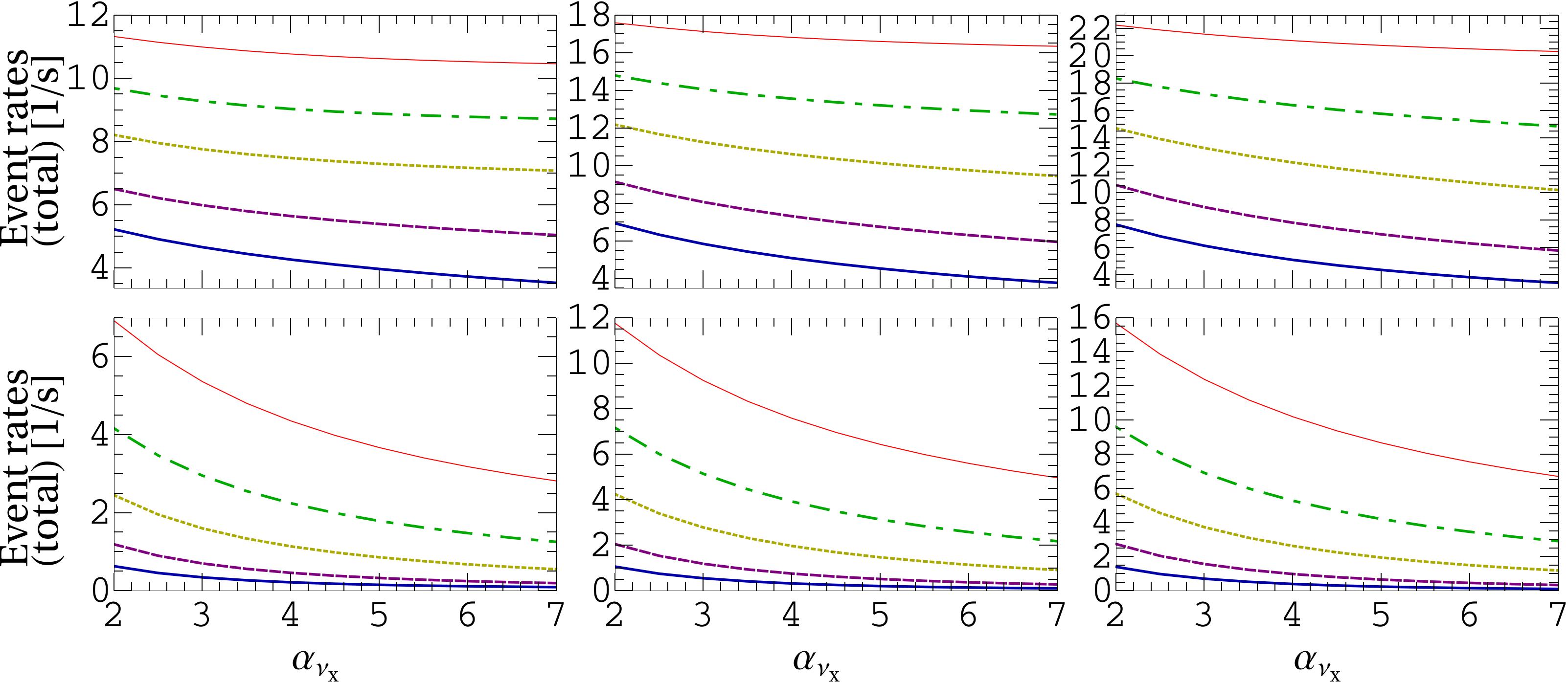}	
	\caption{One- (upper) and two-neutron (lower row) total (charged + neutral current) event rates as a function of primary non-electron neutrino pinching parameter $\alpha_{\nu_x}$, including $\nu-\nu$ interactions, the MSW effect and decoherence.  The case is of $L_{\nu_x} = 2 L_{\nu_e} \ (L_{\bar{\nu}_e} = L_{\nu_e} = 10^{51}\ \mathrm{erg\ s^{-1}})$ in IMH (left), NMH with large $\theta_{13}$ (middle) and NMH with small $\theta_{13}$ (right). Different lines correspond to different non-electron-type primary neutrino average energies (from bottom to top: $13, 15, 18, 21$ and $25\ \mathrm{MeV}$). We assume 100 \% detection efficiency, $L_{\nu}^{\mathrm{TOT}} = 10^{52}\ \mathrm{erg\ s^{-1}}$, distance to the supernova 10 kpc and target mass 1 kton of ${}^{208}$Pb).}
	\label{fig:resALLneqL}
\end{figure*}

Results for $1n$- and $2n$-event rates, as a function of non-electron-type primary neutrino pinching parameter $\alpha_{\nu_x}$, are given in figure~\ref{fig:resALLeqL} for equal luminosities and in figure~\ref{fig:resALLneqL} for $L_{\nu_x} = 2 L_{\nu_e}$, with different non-electron-type primary neutrino average energies $\langle E^0_{\nu_x} \rangle$. These results show how the variation of the parameters describing the primary non-electron type neutrino energy spectrum affects the observable event rates. The results for the normal mass hierarchy with large $\theta_{13}$ are degenerate with inverted mass hierarchy. This is because the low energy split present in the former case is not visible due to the high neutron emission thresholds in lead. 
It is for the same reason that the prompt $\nu_e$ burst occurring at the very early stages of a possible future core-collapse supernova explosion will not be observable in such a detector if the primary electron neutrino average energies are as low as $10\ \mathrm{MeV}$. 
Note that, in the absence of collective effects, the results of IMH would be the same as for the NMH with small $\theta_{13}$ shown here. It is clear that the
1$n$- and 2$n$-events alone will not allow the identification of all the degenerate physical scenarios depending on either the primary flux parameters ($\langle E^0_{\nu_x} \rangle$, $\alpha_{\nu_x}$, neutrino luminosities), or neutrino properties (mass hierarchy and the value of $\theta_{13}$). The neutrino luminosity, that gives the normalization factor of the events shown, can be 
obtained from the measurement of neutral current events in other detectors, as discussed above. 
From figures~\ref{fig:resALLeqL} and~\ref{fig:resALLneqL} it can be seen that $1n$- and $2n$-events provide complementary information: the $1n$-events are more sensitive to pinching with smaller values of $\langle E^0_{\nu_x} \rangle$, while the $2n$-ones are sensitive to high average energy values. 

As far as number of events are concerned, the range of events expected during the accretion phase  are given in table~\ref{tab:eventsACC} in different physical scenarios. For this case a time duration of $300\ \mathrm{ms}$ is assumed and a total luminosity of $18\times10^{52}\ \mathrm{erg\ s^{-1}}$.\footnote{According to e.g.~\cite{Fischer:2009af} for 18 solar mass iron-core progenitor the duration of accretion phase is about $350\ \mathrm{ms}$ and the electron-type neutrino luminosities on average about $0.4\times10^{53}\ \mathrm{erg\ s^{-1}}$(about half of this for non-electron-type).} Therefore, the numbers are obtained by multiplying the events  given in figure~\ref{fig:resALLeqL} by a factor $5.4$ s. It is to be noticed that the collective effects can be suppressed by the background matter during the accretion phase~\cite{Chakraborty:2011gd}. Even if this finding is consolidated, our results remain useful since IMH would be then degenerate with NMH and small $\theta_{13}$ and in NMH collective effects are not present in the neutrino channel. Table~\ref{tab:eventsCOOL} presents the total number of events for the cooling phase using results in figures \ref{fig:resALLeqL} and \ref{fig:resALLneqL} multiplied by a duration time of 10 s. In table~\ref{tab:eventsEqL} the total numbers of events for the whole explosion are given assuming equal luminosities throughout the neutrino emission with the total time-integrated neutrino luminosity fixed at the typical value of $3\times 10^{53}\ \mathrm{erg}$\footnote{These numbers are obtained by multiplying the results in figure \ref{fig:resALLeqL} by a factor of 30 s to obtain the given total time integrated luminosity.}. 

\begin{table*}[t]
	\begin{center}
		\renewcommand{\arraystretch}{1.2}
		\begin{tabular}{c|c|cc|cc|c}
			\hline
			$\langle E^0_{\nu_x} \rangle\ \mathrm{[MeV]}$ & 13 & \multicolumn{4}{c|}{18} & 25 \\
			\hline
			\multirow{2}{*}{MH (and $\theta_{13}$)} & NMH & \multicolumn{2}{c|}{\multirow{2}{*}{IMH}} & \multicolumn{2}{c|}{NMH} & \multirow{2}{*}{IMH} \\
			& small $\theta_{13}$ & & & \multicolumn{2}{c|}{small $\theta_{13}$} & \\
			\hline
			$\alpha_{\nu_x}$ & 7 & 2 & 7 & 2 & 7 & 2 \\
			\hline
			$N_{1n}$ & 16 & 70 & 51 & 54 & 41 & 103 \\
			$N_{2n}$ & $< 3$ & 27 & 5 & 19 & 4 & 70 \\
			\hline
			neutrons emitted  & $\sim 16$ & 124 & 61 & 92 & 49 & 243 \\
			\hline
		\end{tabular}
	\end{center}
\caption{Accretion phase: Total numbers of one- and two-neutron events, $N_{1n}$ and $N_{2n}$, respectively, and total number of neutrons emitted $(N_{1n} + 2 N_{2n})$, depending on the value of the pinching parameter $\alpha_{\nu_x}$ and neutrino properties, including $\nu-\nu$ interactions, the MSW effect and decoherence. Here is  assumed equal neutrino luminosities and 100 \% detection efficiency. The total (sum of all flavors) time-integrated luminosity is taken to be $5.4\times10^{52}\ \mathrm{erg}$ (see text for details), distance to the supernova 10 kpc and target mass 1 kton of ${}^{208}$Pb. The column with 13 (25) $\ \mathrm{MeV}$ represents the scenario with smallest (largest) amount of events. In addition, the range of possible event numbers are shown for the intermediate case with $18\ \mathrm{MeV}$. Notice that the results for NMH and large $\theta_{13}$ are degenerate with ones in IMH shown here.}
	\label{tab:eventsACC}
\end{table*}



\begin{table*}[t]
	\begin{center}
		\renewcommand{\arraystretch}{1.2}
		\begin{tabular}{c|cc|cc|cc}
			\hline
			\multirow{2}{*}{MH (and $\theta_{13}$)} & \multicolumn{2}{c|}{\multirow{2}{*}{IMH}} & \multicolumn{2}{c|}{NMH} & \multicolumn{2}{c}{NMH} \\
			& & & \multicolumn{2}{c|}{large $\theta_{13}$} & \multicolumn{2}{c}{small $\theta_{13}$} \\
			\hline
			$\alpha_{\nu_x}$ & 2 & 7 & 2 & 7 & 2 & 7 \\
			\hline
			$N_{1n}$ & 82(130) & 70(95) & 122(130) & 95(95) & 150(100) & 102(75) \\
			$N_{2n}$ & 25(50) & 6(10) & 45(50) & 10(10) & 58(35) & 12(8) \\
			\hline
			neutrons emitted & 132(230) & 82(115) & 212(230) & 115(115) & 266(162) & 126(91) \\
			\hline
		\end{tabular}
	\end{center}
\caption{Cooling phase: Total numbers of one- and two-neutron events, $N_{1n}$ and $N_{2n}$, respectively, and total number of neutrons emitted, depending on the pinching parameter $\alpha_{\nu_x}$ and neutrino properties, including $\nu-\nu$ interactions, the MSW effect and decoherence. Here is assumed that
$L_{\nu_x}=2L_{\nu_e}$ (equal luminosities in parenthesis) with the total time integrated luminosity $10^{53}\ \mathrm{erg}$ and $\langle E^0_{\nu_x} \rangle = 18\ \mathrm{[MeV]}$. 
The calculations are obtained using 100 \% detection efficiency, a distance to the supernova 10 kpc and a target mass 1 kton of ${}^{208}$Pb. 
The column with 13 (25) $\ \mathrm{MeV}$ represents the scenario with smallest (largest) amount of events.}
	\label{tab:eventsCOOL}
\end{table*}

\begin{table*}[t]
	\begin{center}
		\renewcommand{\arraystretch}{1.2}
		\begin{tabular}{c|c|cc|cc|c}
			\hline
			$\langle E^0_{\nu_x} \rangle\ \mathrm{[MeV]}$ & 13 & \multicolumn{4}{c|}{18} & 25 \\
			\hline
			\multirow{2}{*}{MH (and $\theta_{13}$)} & NMH & \multicolumn{2}{c|}{\multirow{2}{*}{IMH}} & \multicolumn{2}{c|}{NMH} & \multirow{2}{*}{IMH} \\
			& small $\theta_{13}$ & & & \multicolumn{2}{c|}{small $\theta_{13}$} & \\
			\hline
			$\alpha_{\nu_x}$ & 7 & 2 & 7 & 2 & 7 & 2 \\
			\hline
			$N_{1n}$ & 90 & 390 & 285 & 300 & 225 & 570 \\
			$N_{2n}$ & $< 3$ & 150 & 30 & 105 & 24 & 390 \\
			\hline
			neutrons emitted  & $\sim 90$ & 690 & 345 & 510 & 273 & 1350 \\
			\hline
		\end{tabular}
	\end{center}
\caption{Total numbers of events during the explosion (assuming 100 \% detection efficiency, distance to the supernova 10 kpc and target mass 1 kton of ${}^{208}$Pb). As in table \ref{tab:eventsACC} but assuming equal neutrino luminosities throughout the whole neutrino emission and the total time integrated luminosity $3\times 10^{53}\ \mathrm{erg}$. 
	}
	\label{tab:eventsEqL}
\end{table*}

\begin{figure*}
\includegraphics[width=1.\textwidth]{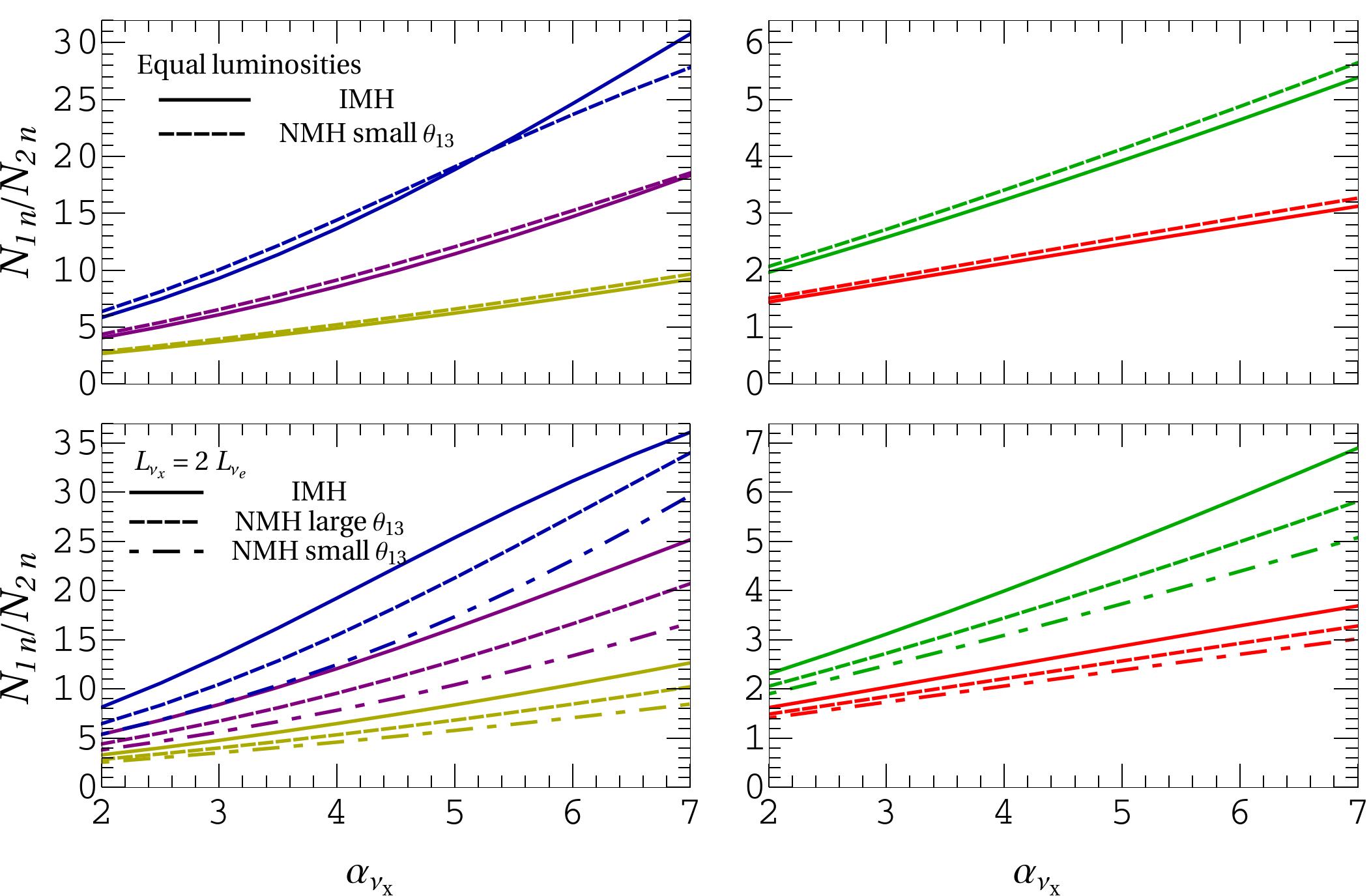}
	\caption{Ratios of one- ($N_{1n}$) to two-neutron ($N_{2n}$) event rates as a function of primary non-electron neutrino pinching parameter $\alpha_{\nu_x}$ with different primary non-electron-type neutrino average energies. Predictions include $\nu-\nu$ interactions, the MSW effect and decoherence. It is assumed 100 \% detection efficiency, distance to the supernova 10 kpc and target mass 1 kton of ${}^{208}$Pb. Upper row: equal luminosities in IMH (solid), NMH with small $\theta_{13}$ (dashed). Lower row: $L_{\nu_x} = 2 L_{\nu_e}\ (L_{\bar{\nu}_e} = L_{\nu_e})$ in IMH (solid), NMH with large $\theta_{13}$ (dashed), NMH with small $\theta_{13}$ (dash-dotted). Left panel: average energies from top to bottom $\langle E^0_{\nu_x} \rangle = 13, 15$ and $18\ \mathrm{MeV}$. Right panel: $\langle E^0_{\nu_x} \rangle = 21$ (higher) and $\langle E^0_{\nu_x} \rangle = 25\ \mathrm{MeV}$ (lower curves).}
	\label{fig:res1n2n}
\end{figure*}

\begin{figure*}[h]
 \subfigure{
 \includegraphics[width=.49\textwidth]{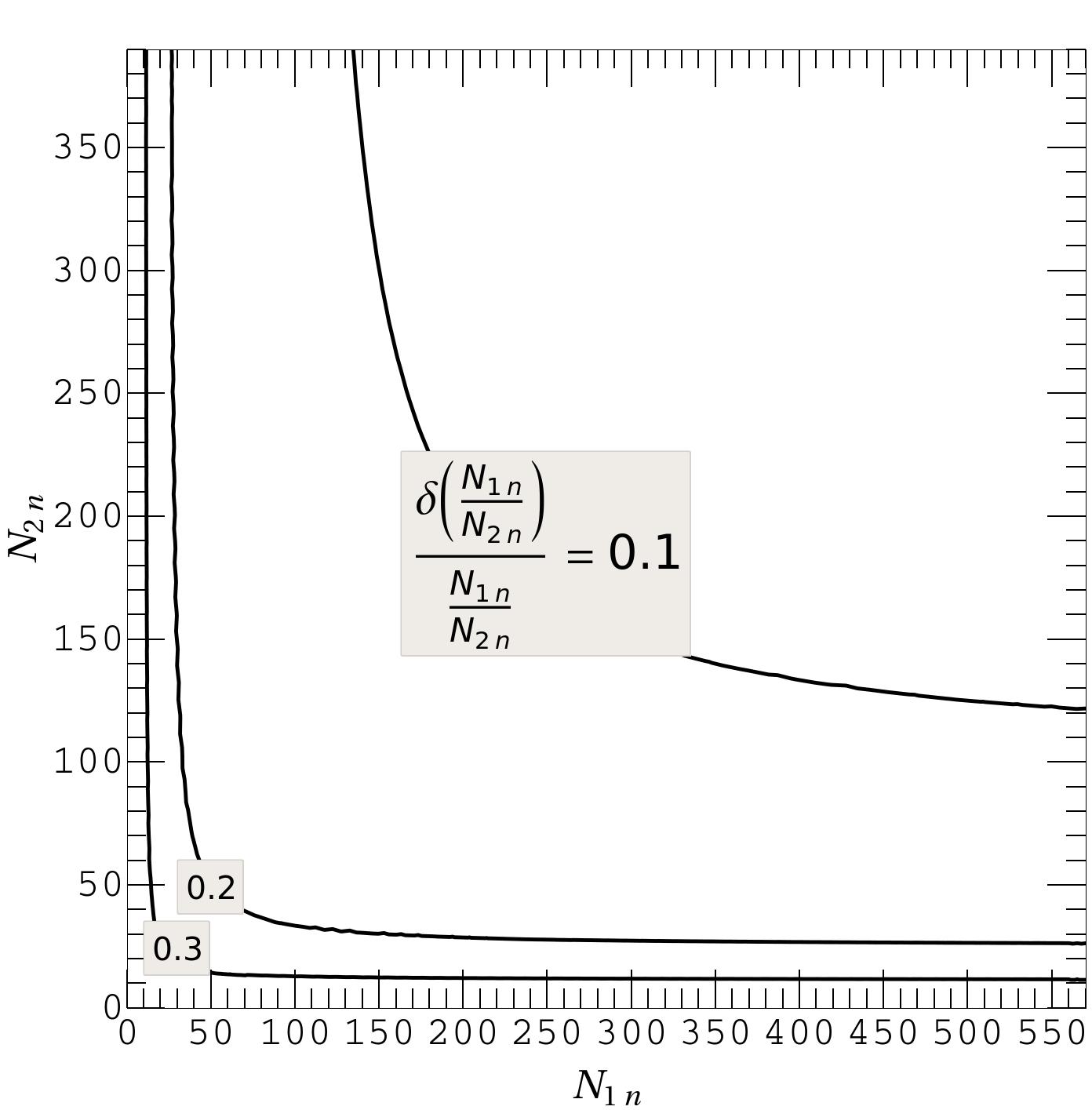}
 \label{fig:ratioErrors1}
}
 \subfigure{
 \includegraphics[width=.49\textwidth]{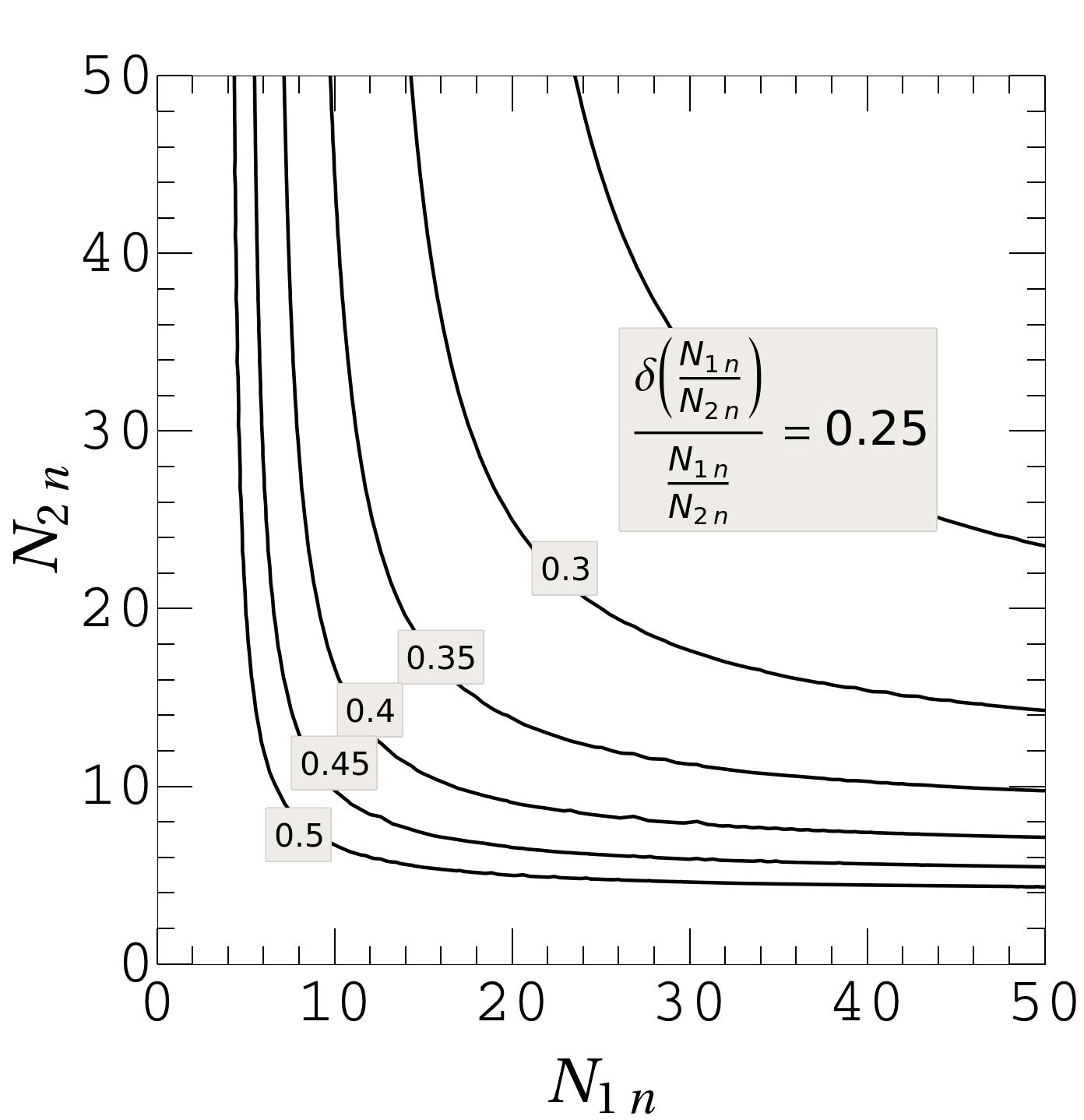}
 \label{fig:ratioErrors2}
 }
\label{fig:ratioErrors}
\caption{Relative statistical error $\frac{\delta\left(N_{1n}/N_{2n}\right)}{N_{1n}/N_{2n}}$ for the ratio $N_{1n}/N_{2n}$ shown in Figure \ref{fig:res1n2n} depending on the numbers of events. Left: The whole range of possible event numbers in our considerations. Right: Zoomed from the left figure showing only for the event numbers $\leq 50$. Above each contour line the statistical error is smaller than the value indicated in the figure. 
For instance, assuming 30 one-neutron and 12 two-neutron events, their ratio is 2.5 and the statistical error about 35\% of the ratio: $N_{1n}/N_{2n} = 2.5 \pm 0.75$.}
\end{figure*}

Since $1n$- and $2n$-events can be well identified in HALO, it is attractive to consider the corresponding ratio (figure~\ref{fig:res1n2n}), which is independent of common normalization factors (such as a distance to the supernova and a size of the detector). 
From figure~\ref{fig:res1n2n} it is evident that these ratios are sensitive to the pinching regardless of the physical scenario i.e. luminosities, neutrino mass hierarchy and the value of $\theta_{13}$.\footnote{Notice the curves in the case of equal luminosities are within the ones of IMH (solid) and NMH with small $\theta_{13}$ (dashed) in $L_{\nu_x} = 2 L_{\nu_e}$ case.} Since in HALO, if a supernova explodes, one measures a given ratio, figure~\ref{fig:res1n2n} shows that the measured ratio would allow to identify different degenerate combinations of non-electron-type primary neutrino average energies and pinching parameters (even without a precise knowledge of the above mentioned physical scenarios and common flux parameters). By knowing (or assuming) the non-electron-type primary neutrino average energy\footnote{In the cases we have been considering the final electron neutrino fluxes are always close to the non-electron-type neutrino primary fluxes (see figure~\ref{fig:Fe}). More precisely, in the equal luminosity case, both in inverted and normal mass hierarchy with large $\theta_{13}$, the final electron neutrino energy spectrum is the same as the primary energy spectrum of non-electron type neutrinos due to the complete flavor conversion (in IMH after the low energy split). In normal mass hierarchy with small $\theta_{13}$ there are less electron neutrinos at higher energies (relevant from the detection point of view) with respect to the primary non-electron type neutrinos. In the $L_{\nu_x} = 2 L_{\nu_e}$ case the average energy of the observable electron neutrinos is somewhat smaller than that of the primary non-electron type neutrinos in all of the studied cases. Notice that the figure~\ref{fig:Fe} corresponds only to the case with $\langle E^0_{\nu_x} \rangle = 18\ \mathrm{MeV}$ while the exact relation between the observable $\nu_e$ and the primary $\nu_x$ spectrum obviously depends on the exact value of $\langle E^0_{\nu_x} \rangle$.}, it would be possible to give tight constraints on pinching. The more favorable case is the one with equal luminosities since the degenerate combination of average energies and pinching parameters would have only a slight dependence on the unknown neutrino properties (figure~\ref{fig:res1n2n}). 
Note that, while in very few cases the event number is small, in the majority of cases statistics is sufficiently large so that the statistical error is not a limiting factor. This is shown in figure \ref{fig:ratioErrors2} where the area above each curve gives the relative statistical errors associated with the event ratios of figure \ref{fig:res1n2n}. 

\begin{figure*}
\includegraphics[width=1.\textwidth]{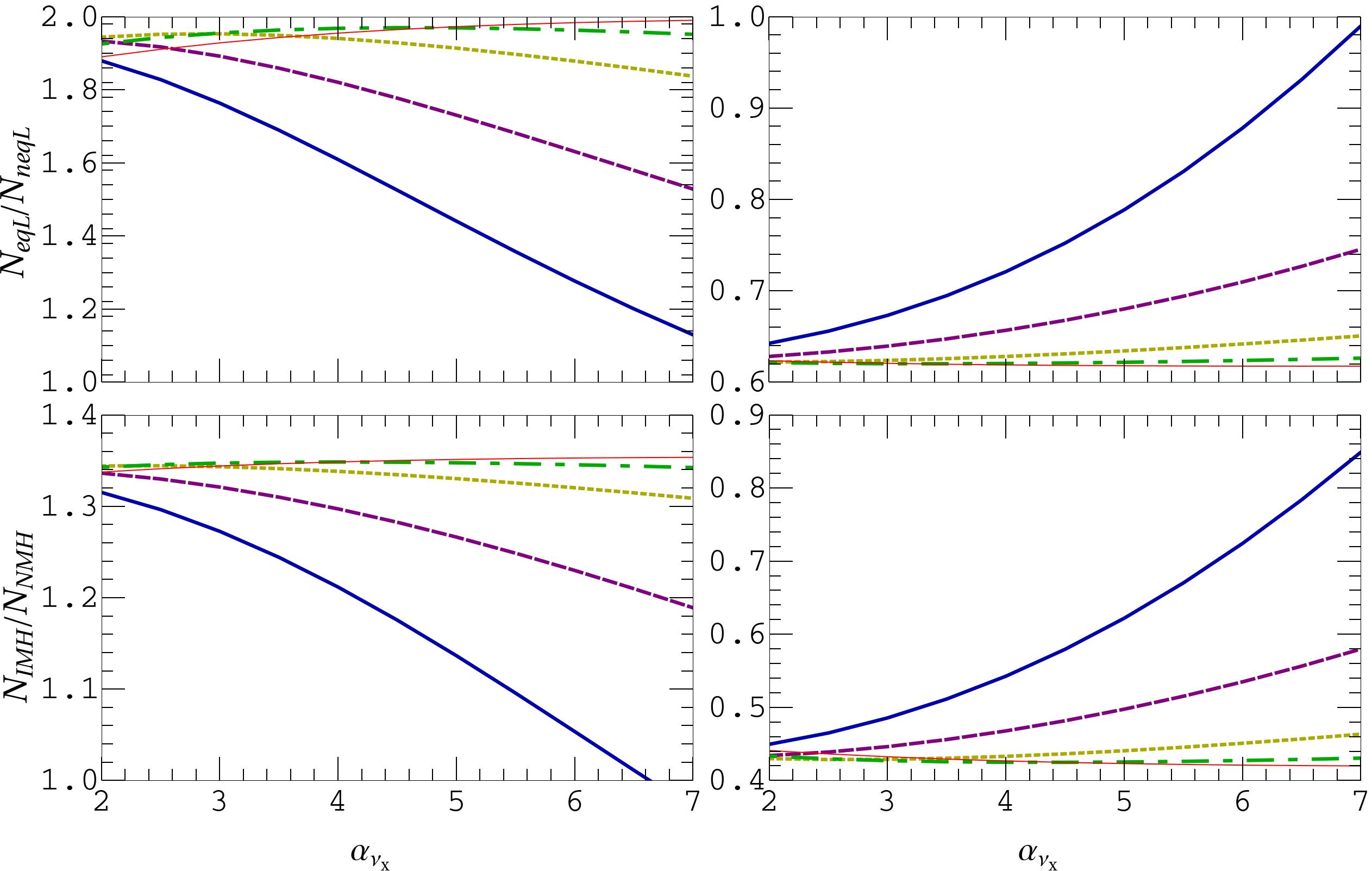}
	\caption{Ratios of two-neutron event rates as a function of primary non-electron-type neutrino pinching parameter $\alpha_{\nu_x}$ with different primary non-electron-type neutrino average energies (as in figure~\ref{fig:resALLeqL}). Upper figures: results of equal luminosities to $L_{\nu_x} = 2 L_{\nu_e}$ in IMH (left) and NMH with small $\theta_{13}$ (right). Lower figures: results in IMH to NMH with small $\theta_{13}$ in the case of equal luminosities (left) and $L_{\nu_x} = 2 L_{\nu_e}$ (right).}
	\label{fig:resRATIOS}
\end{figure*}

In figure~\ref{fig:resRATIOS} we give ratios of $2n$-events to determine which precision should be achieved in measuring the numbers of events to be able to distinguish different physical scenarios. To quantify this, First, we assume the neutrino mass hierarchy to be known (inverted or normal with small $\theta_{13}$) and present the ratios of the $2n$-events for the case with equal luminosities over the one with $L_{\nu_x} = 2 L_{\nu_e}$. Second, we fix the luminosity (either equal or $L_{\nu_x} = 2 L_{\nu_e}$) and show the ratios of the $2n$-events with IMH over NMH with small $\theta_{13}$. While these ratios are obviously not observable, we employ them to give an idea of the required precision to be able to identify either different neutrino properties or supernova emission cases. From  figure~\ref{fig:resRATIOS} one can clearly see that the difference between the results in different scenarios range from 30 \% up to a factor of 2\footnote{Note that, while we use a difference between $L_{\nu_x}$ and $ L_{\nu_e}$ as large as a factor of 2, similar results are obtained if such a difference is as low as 30 \%.}. 
The largest difference is due to the presence of the high energy split in the neutrino fluxes. Note that a detector capable of measuring the neutrino spectra like a scintillator or a liquid argon supernova observatory can actually observe the high energy split that would further restrict
the range of possible primary neutrino flux parameters. 
Without such an information, HALO alone can give an indication on the presence of energy equipartition or
on $L_{\nu_x}$ being larger than $ L_{\nu_e}$.
 
It is clear that to be able to extract the most from future observations, 
a precise measurement of neutrino-lead cross sections is called for. 
This could be realized either at a low energy beta-beam facility proposed by 
\cite{Volpe:2003fi}, or nearby one of the future intense Spallation Sources 
\cite{Avignone:2003ep,Lazauskas:2010rh}. Indeed a broad physics program can be 
realized if a low energy neutrino facility covering from a few MeV to 100$\ \mathrm{MeV}$ energy range becomes available in the future, including cross section measurements, fundamental interactions tests (neutrino-nucleus coherent scattering, the measurement of  the Weinberg angle at low momentum transfer, a CVC test, neutrino oscillations to sterile) and measurements of interest for core-collapse supernova physics (see \cite{Volpe:2003fi,Avignone:2003ep,Volpe:2006in} and references therein).

\begin{figure*}[h]
\centering
 \subfigure{
 \includegraphics[width=.47\textwidth]{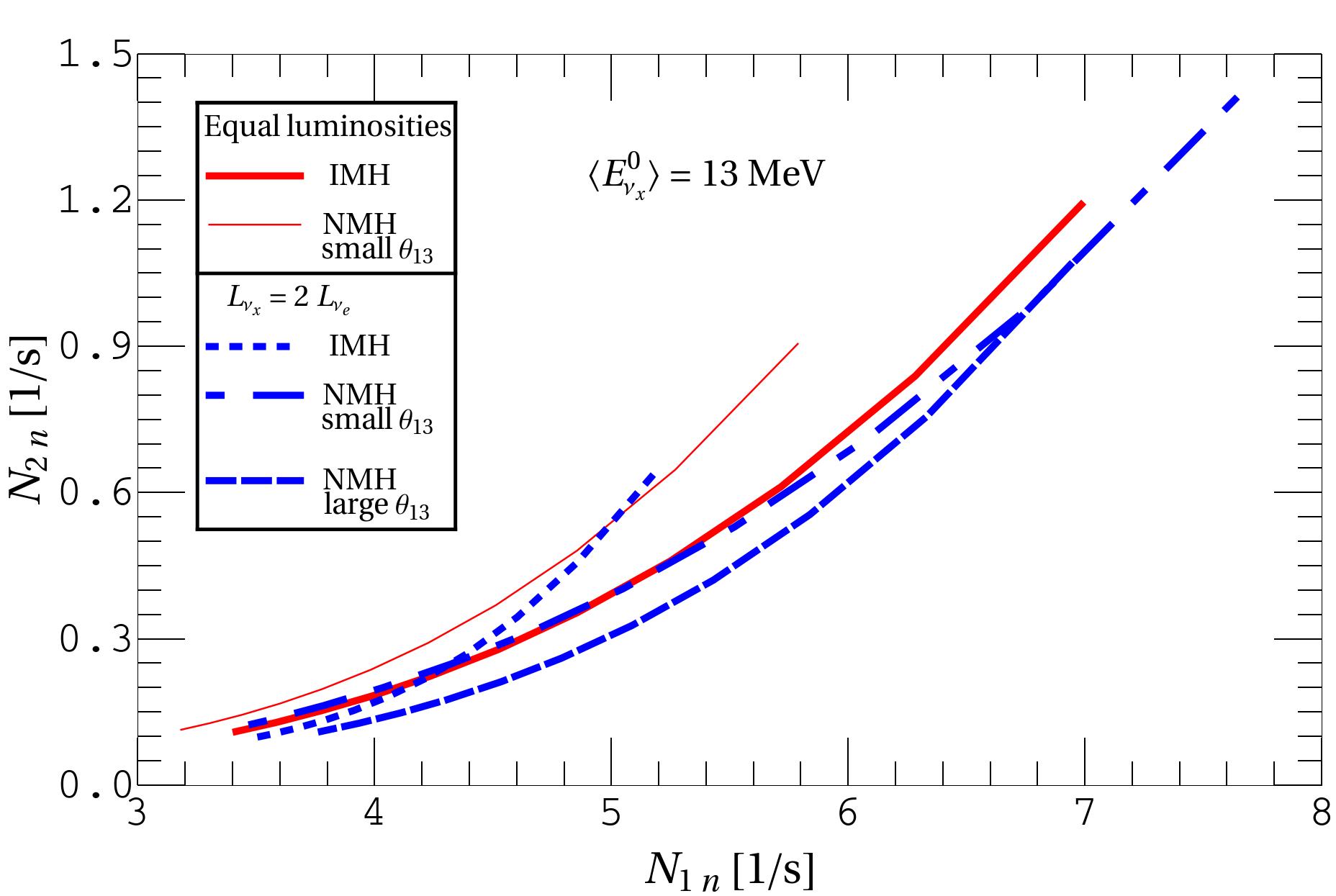}
 \label{fig:sum13}
}
 \subfigure{
 \includegraphics[width=.47\textwidth]{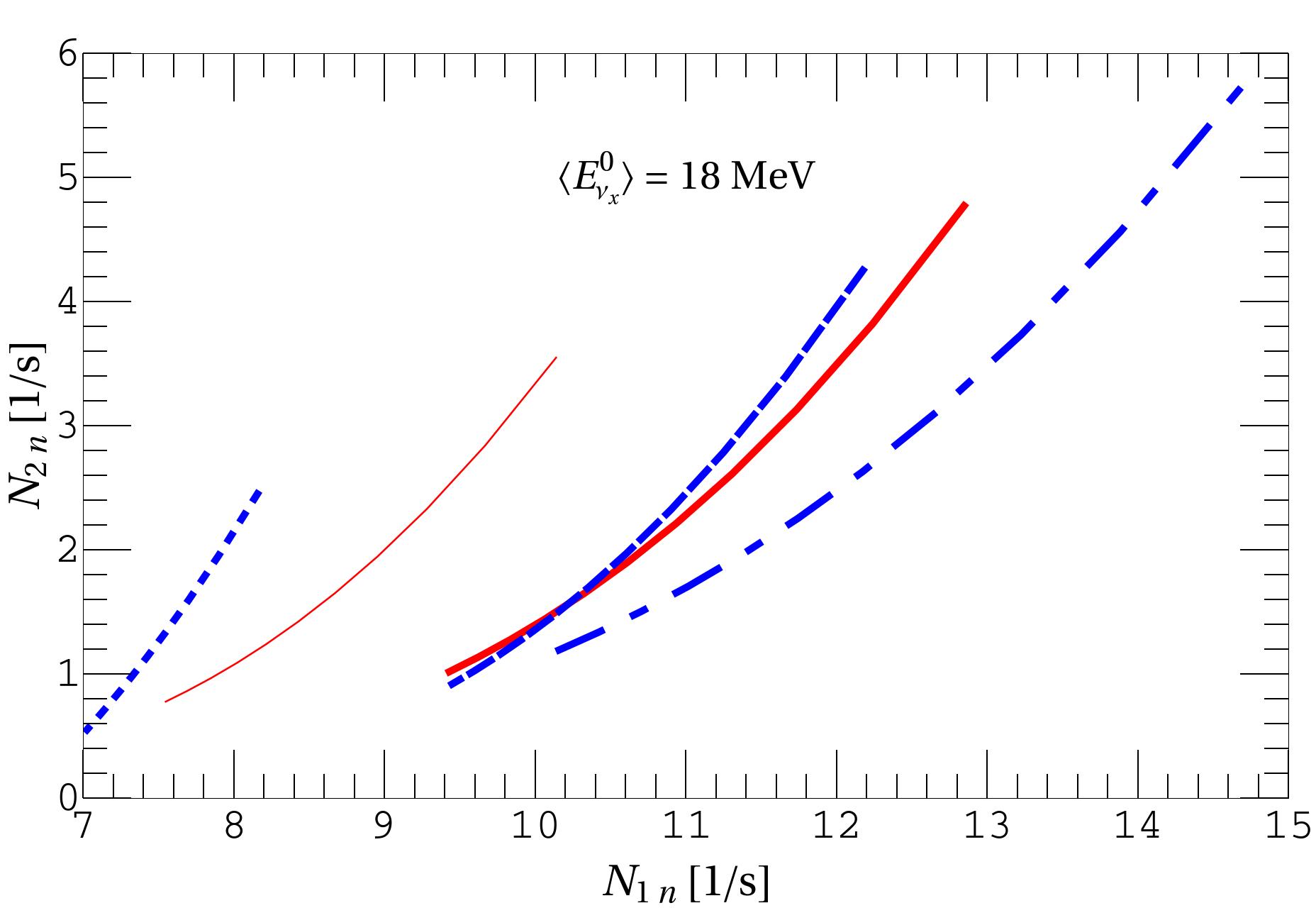}
 \label{fig:sum18}
 }
  \subfigure{
 \includegraphics[width=.47\textwidth]{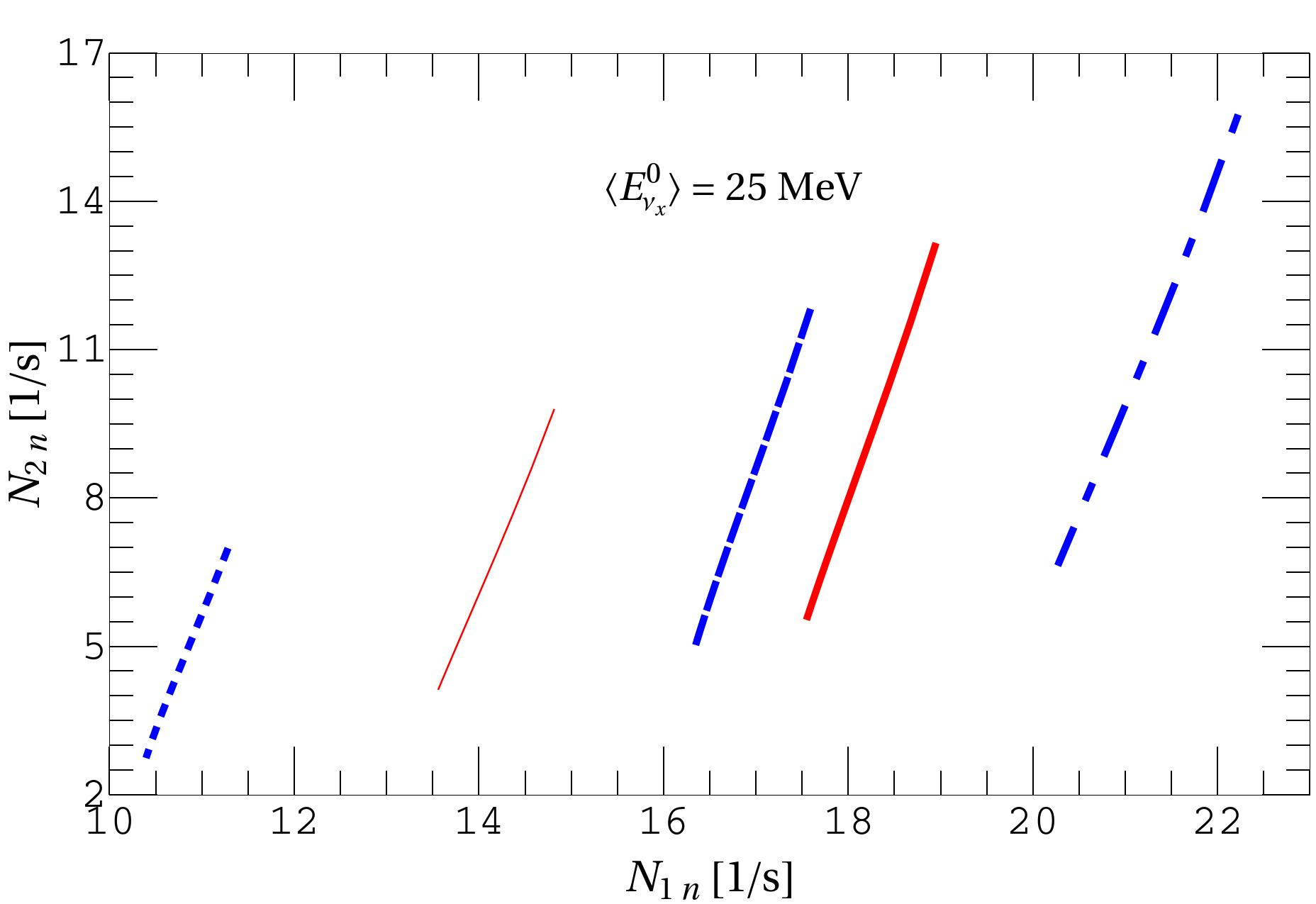}
 \label{fig:sum25}
 }
	\caption{(Color online) One- and two-neutron emission event rates 
	with different values of the primary non-electron neutrino pinching parameter $\alpha_{\nu_x}$: at the top of each curve $\alpha_{\nu_x} = 2$, at the bottom $\alpha_{\nu_x} = 7$. Results include $\nu-\nu$ interactions, the MSW effect and decoherence. The primary non-electron neutrino average energies are fixed: $\langle E^0_{\nu_x} \rangle = 13\ \mathrm{MeV}$ (top left), $\langle E^0_{\nu_x} \rangle = 18\ \mathrm{MeV}$ (top right) and $\langle E^0_{\nu_x} \rangle = 25\ \mathrm{MeV}$ (bottom figure). Solid lines are for equal luminosities (thick IMH, thin NMH with small $\theta_{13}$), others for $L_{\nu_x} = 2 L_{\nu_e}$: dotted IMH, dashed NMH with large $\theta_{13}$ and dash-dotted NMH with small $\theta_{13}$. 
The total luminosity is taken to be $10^{52}\ \mathrm{erg\ s^{-1}}$, distance to the supernova 10 kpc, a target mass 1 kton of ${}^{208}$Pb and 100 \% detection efficiency is assumed. Notice that the results in the case of equal luminosities in NMH with large $\theta_{13}$ are the same as the ones in IMH.}
	\label{fig:resALL}
\end{figure*}

We have summarized our results in figure~\ref{fig:resALL}, where all the $1n$-and $2n$-event rates are plotted 
in different physical scenarios with a fixed non-electron-type primary neutrino average energy. In these figures a given measurement would correspond to just one point. It can be seen that if the primary $\langle E^0_{\nu_e} \rangle$ and $\langle E^0_{\nu_x} \rangle$ differ by little, the curves have strong overlap, making more difficult to see any evidence of flavor conversion. Obviously, more the primary average energies differ, easier it is to identify a given physical scenario. In the extreme case where the primary $\langle E^0_{\nu_e} \rangle$ and $\langle E^0_{\nu_x} \rangle$ differ as much as $15\ \mathrm{MeV}$ most of the scenarios are very likely to be distinguished. 

\begin{figure*}
\includegraphics[width=1.\textwidth]{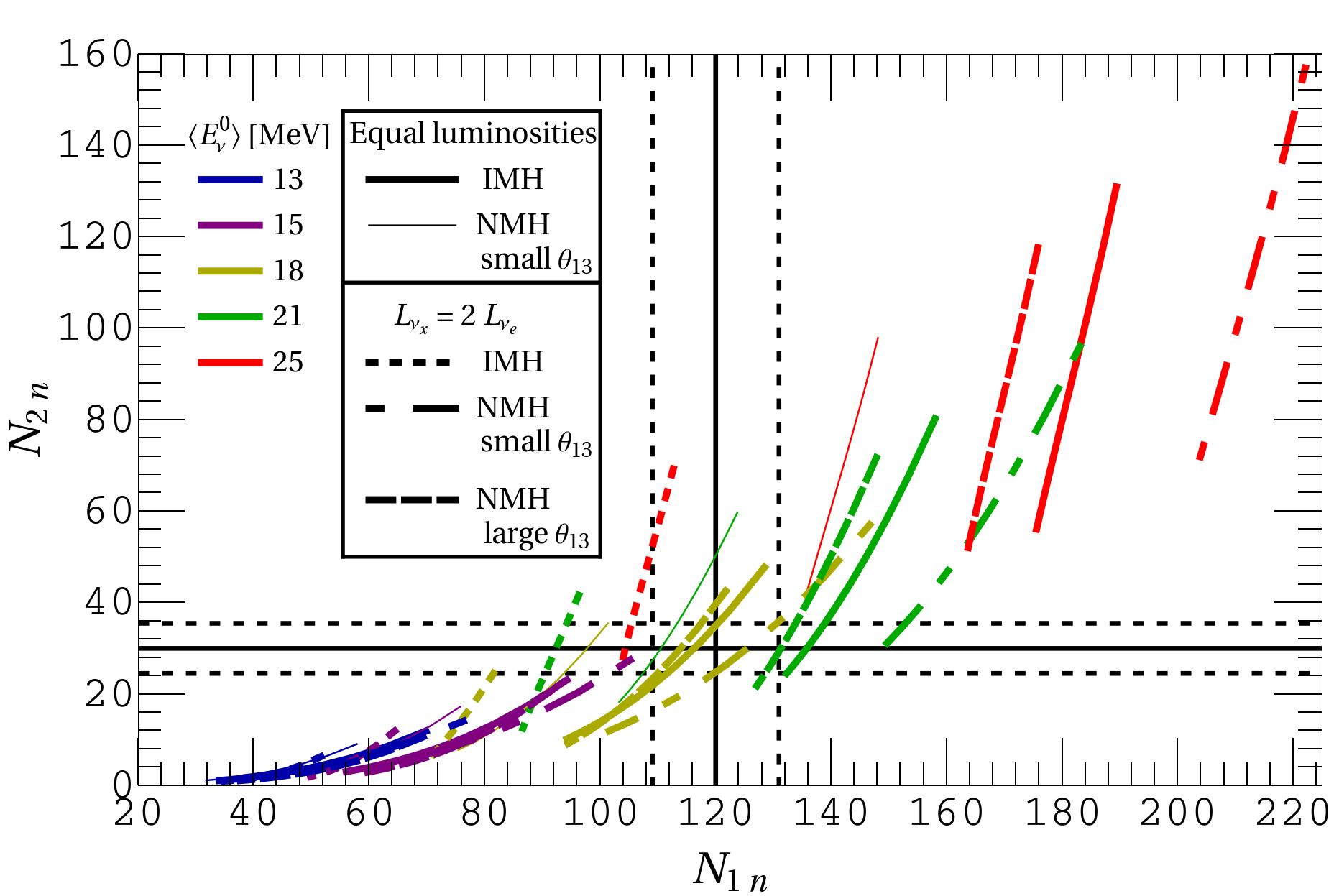}
	\caption{(Color online) One- and two-neutron emission event rates 
	with different values of the primary non-electron neutrino pinching parameter $\alpha_{\nu_x}$: at the top of each curve $\alpha_{\nu_x} = 2$, at the bottom $\alpha_{\nu_x} = 7$. Results include $\nu-\nu$ interactions, the MSW effect and decoherence. Different colors correspond to different non-electron type primary average energies (as in figure \ref{fig:resALLeqL}), solid lines are for equal luminosities (thick IMH, thin NMH with small $\theta_{13}$), others for $L_{\nu_x} = 2 L_{\nu_e}$: dotted IMH, dashed NMH with large $\theta_{13}$ and dash-dotted NMH with small $\theta_{13}$. 
The total time-integrated luminosity is taken to be $10^{53}\ \mathrm{erg}$ ($\sim$ total events during the cooling phase), distance to the supernova 10 kpc, a target mass 1 kton of ${}^{208}$Pb and 100 \% detection efficiency is assumed. Notice that the results in the case of equal luminosities in NMH with large $\theta_{13}$ are the same as the ones in IMH. In addition is shown an example measurement case with 120 one-neutron and 30 two-neutron events (solid black lines) together with estimated statistical errors 
(dashed black lines).}
	\label{fig:resALL2}
\end{figure*}

The whole set of our results for the one- and two-neutron number of events associated with a typical cooling phase are shown in figure \ref{fig:resALL2}. The straight lines correspond to an example where 120 one-neutron events and 30 two-neutron events are measured during the explosion, with the associated statistical error.  In this example case, one can see that, while from the point of view of neutrino properties (mass hierarchy and third neutrino mixing angle) all scenarios are possible, rather tight constraints on the primary flux
parameters -- average energy and pinching parameter -- can be obtained. Notice that, if the $1n$- and $2n$-event numbers are 210 and 140, respectively (corresponding to $\langle E^0_{\nu_x} \rangle \approx 25\ \mathrm{MeV}$ and $\alpha_{\nu_x} \approx 2-4$), one can obtain also a clear indication on the mass hierarchy, the value of $\theta_{13}$ and the luminosity case : the most favorable is $L_{\nu_x} = 2 L_{\nu_e}$ and NMH with small $\theta_{13}$). It can be also seen that the indication on the presence or absence of the energy equipartition can be gained with smaller $\langle E^0_{\nu_x} \rangle$ if other information -- mass hierarchy and/or $\theta_{13}$ -- is known from other experiments, as mentioned above.
 
In the present work we propagate neutrino probabilities instead of amplitudes. 
Thus, phase effects coming from the shock wave and turbulence effects are not included. 
Nevertheless, our results represent a good reference of what can be measured. 
First of all, they give an adequate description at early (late) times, when 
the shock wave has not yet reached (left) the MSW region, that will be well identified 
given the good time resolution of the HALO detector. 
Shock waves and turbulence only impact the signal in the case of normal hierarchy and large third neutrino
mixing angle when the shock wave has reached the MSW region. For such a case 
one can consider that the neutrino probabilities average approximately to 1/2, giving 
a result that is intermediate between the adiabatic and non-adiabatic cases we have 
been calculating. This is also a good approximation for turbulence effects as far as 
the matter density fluctuations keep small enough since large matter density fluctuations
might break the HL factorization employed in the present work
\cite{Kneller:2010ky,Kneller:2010sc}. Obviously, to have a detailed description of neutrino 
flavor conversion and an accurate prediction of the event rates in all cases, the inclusion 
of shock waves and turbulence should be implemented in a future calculation. 
\section{Conclusions}
\noindent
We have presented new predictions of the expected neutrino events from iron core-collapse 
supernovae in a lead-based observatory. Such a detector, Helium and Lead Observatory 
(HALO), is currently under construction at SNOLAB. Our predictions are based upon the 
formalism of factorized dynamics. In particular our calculations include collective 
flavor conversion and the MSW effects, as well as decoherence coming from the propagation 
up to the Earth. Determining the luminosities, primary average energies and pinching 
parameters allows an identification of the primary neutrino fluxes at the neutrinosphere, 
which furnishes valuable information to test our understanding of the neutrino transport in 
the supernova.

We have shown that  in case neutrino flavor transformations occur in supernovae the measurement of $1n$- and $2n$-event rates as well as of their ratio is particularly sensitive to the non-electron neutrino primary average energy and pinching parameter and that interesting information can be extracted. Indeed, using information from other detectors, the combination of $1n$- and $2n$-event rates should allow to identify degenerate solutions of average energies and pinching values. Moreover, from the ratio of these events, HALO alone can be used to give constraints on these parameters. If the non-electron type primary average energy is identified using other detectors one can get even better constraints on the primary 
pinching parameter. 

The actual equipartition of the gravitational energy among the neutrino flavors 
is also an unknown. If the muon or tau neutrino 
luminosities are larger than the electron neutrino ones 
- as some supernova simulations 
of the cooling phase seem to indicate - 
a high energy split might appear, producing different event rates compared to the case where the assumption of equipartition of energy among the neutrinos is made. From the combination of $1n$- and $2n$-event rate measurement one can give an indication on the presence or absence of this equipartition depending on the actual values of the flux parameters. More the primary neutrino average energies differ and if other information (mass hierarchy and/or $\theta_{13}$) is known, better indications can be given.

The present work emphasizes the interest of having more information on the characteristics of the high-energy component 
of the neutrino distributions at the neutrinosphere from future supernova simulations:
the flux tail is essential to precisely predict the events in such a detector.
Examples of this type are not only 
HALO that has been the object of the present study but also scintillator detectors using carbon or a Cherenkov detector having oxygen. All these have the interesting characteristics of being sensitive to specific features of the high energy tail of the neutrino fluxes in a signal from a future supernova explosion. 

Our specific study for a lead-based observatory furnishes a good example of how, having a network of detectors with different energy thresholds, constitute a unique tool to probe different components of the neutrino fluxes from a core-collapse supernova explosion and to unravel interesting information on the neutrino emission and on neutrino properties. The capability of extracting as much information as possible from a future supernova neutrino signal in HALO will obviously rely upon the
precision with which measurements will be made. This involves, in particular, a
precise knowledge of the associated cross sections that could be at reach with a low energy beta-beam facility or nearby future Spallation Sources. 

\acknowledgments{
We thank Nils Paar for providing us with the neutrino-lead cross sections
within the relativistic Random-Phase approximation, as in ref.\cite{Paar:2007fi}, 
and Clarence Virtue for useful discussions concerning the characteristics of the HALO observatory.
}

\clearpage

\bibliographystyle{JHEP}
\bibliography{refs}

\end{document}